\documentclass[superscriptaddress,twocolumn,nopreprintnumbers,floatfix,nofootinbib,trackchanges,table]{aastex631}
\pdfoutput=1
\usepackage[utf8]{inputenc}
\usepackage{xspace}
\usepackage{acronym}
\usepackage{booktabs}
\usepackage{amsmath,amssymb,graphicx}
\usepackage{color}
\usepackage{xcolor}
\usepackage{tabularx}
\usepackage{multirow}
\usepackage{makecell}
\usepackage{array}

\usepackage{longtable}

\newcolumntype{Y}{>{\centering\arraybackslash}X}
\newcolumntype{Z}{>{\raggedright\arraybackslash}X}
\newcolumntype{K}{>{\raggedleft\arraybackslash}X}
\newcolumntype{U}{>{\hsize=1.01\hsize}Y}
\newcolumntype{V}{>{\hsize=1.2\hsize}Y}
\newcolumntype{W}{>{\hsize=0.71\hsize}Y}

\usepackage{scrextend}
\usepackage{textgreek}
\usepackage[caption=false]{subfig}

\usepackage{makerobust}

\renewcommand{\today}{\number\day\space\ifcase\month\or
  January\or February\or March\or April\or May\or June\or
  July\or August\or September\or October\or November\or December\fi
  \space\number\year}

\acrodef{LIGO}[LIGO]{Laser Interferometer Gravitational-Wave Observatory}
\acrodef{LHO}[LHO]{\ac{LIGO}-Hanford Observatory}
\acrodef{LLO}[LLO]{\ac{LIGO}-Livingston Observatory}
\acrodef{KAGRA}[KAGRA]{Kamioka Gravitational Wave Detector}
\acrodef{GEO}[GEO]{GEO\,600 Gravitational Wave Detector}
\acrodef{aLIGO}{Advanced \ac{LIGO}}
\acrodef{AdV}{Advanced \ac{Virgo}}
\acrodef{Virgo}{Virgo}\acused{Virgo}

\acrodef{LSC}[LSC]{\ac{LIGO} Scientific Collaboration}
\acrodef{LVC}[LVC]{\ac{LSC} and Virgo Collaboration}
\acrodef{LVK}[LVK]{\ac{LSC}, Virgo Collaboration, and \ac{KAGRA} Collaboration}

\acrodef{O1}[O1]{first observing run}
\acrodef{O2}[O2]{second observing run}
\acrodef{O3}[O3]{third observing run}
\acrodef{O3a}[O3a]{first half of the third observing run}
\acrodef{O3b}[O3b]{second half of the third observing run}
\acrodef{O3GK}[O3GK]{observing run}
\acrodef{O4}[O4]{fourth observing run}
\acrodef{O4a}[O4a]{first part of the fourth observing run}
\acrodef{O4b}[O4b]{second part of the fourth observing run}
\acrodef{O4c}[O4c]{third part of the fourth observing run}

\acrodef{BH}[BH]{black hole}
\acrodef{BBH}[BBH]{binary black hole}
\acrodef{BNS}[BNS]{binary neutron star}
\acrodef{IMBH}[IMBH]{intermediate-mass black hole}
\acrodef{NS}[NS]{neutron star}
\acrodef{BHNS}[BHNS]{black hole--neutron star binaries}
\acrodef{NSBH}[NSBH]{neutron star--black hole binary}
\acrodef{PBH}[PBH]{primordial black hole binaries}
\acrodef{CBC}[CBC]{compact binary coalescence}

\acrodef{GW}[GW]{gravitational wave}
\acrodef{GWH}[GW]{gravitational-wave}
\acrodef{IFO}[IFO]{interferometer}

\acrodef{CWB}[cWB]{coherent WaveBurst}

\acrodef{SNR}[SNR]{signal-to-noise ratio}
\acrodef{FAR}[FAR]{false alarm rate}
\acrodef{IFAR}[IFAR]{inverse false alarm rate}
\acrodef{FAP}[FAP]{false alarm probability}
\acrodef{PSD}[PSD]{power spectral density}

\acrodef{GR}[GR]{general relativity}
\acrodef{NR}[NR]{numerical relativity}
\acrodef{PN}[PN]{post-Newtonian}
\acrodef{EOB}[EOB]{effective-one-body}
\acrodef{ROM}[ROM]{reduced-order model}
\acrodef{IMR}[IMR]{inspiral--merger--ringdown}

\acrodef{PDF}[PDF]{probability density function}
\acrodef{PE}[PE]{parameter estimation}
\acrodef{CL}[CL]{credible level}

\acrodef{EOS}[EoS]{equation of state}

\acrodef{LAL}[LAL]{LIGO Algorithm Library}

\acrodef{KLD}[KLD]{Kullback--Leibler divergence}
\acrodef{JSD}[JSD]{Jensen--Shannon divergence}

\acrodef{GCN}[GCN]{General Coordinates Network}

\newcommand\gwtc[1][?]{\mbox{GWTC\if#1?\else-#1\fi}}
\newcommand\thisgwtcversionmajor{4}
\newcommand\thisgwtcversionminor{0}
\newcommand\thisgwtcversionfull{\thisgwtcversionmajor.\thisgwtcversionminor}
\newcommand\thisgwtcversion\thisgwtcversionfull

\newcommand{\pastro}{\ensuremath{p_{\mathrm{astro}}}\xspace}

\providecolor{UNREVIEWED}{rgb}{1,0,0}
\providecolor{REVIEWED}{rgb}{0,.5,.5}

\acrodef{FPP}[FPP]{false-positive probability}
\acrodef{HOM}[HOM]{higher-order mode}
\acrodef{JPE}[JPE]{joint parameter estimation}
\acrodef{PO}[PO]{posterior overlap}
\acrodef{SDSS}[SDSS]{Sloan Digital Sky Survey}
\acrodef{SIS}[SIS]{singular isothermal sphere} % for extra lensing acronyms
\definecolor{NOTECOLOR}{rgb}{0.6, 0.2, 0.1}
\definecolor{notecolor}{rgb}{0.1, 0.5, 0.9}
\definecolor{darkgreen}{rgb}{0.1, 0.8, 0.2}
\newcommand{\boldtheta}{\boldsymbol{\theta}} 
\newcommand{\boldphi}{\boldsymbol{\phi}}

\providecolor{UNREVIEWED}{rgb}{1,0,0}
\providecolor{REVIEWED}{rgb}{0,.5,.5}
\newcommand{\GTWCFourTotalNumberPastroPointFive}{128 }

\newcommand{\OFourATotalNumberPastroPointFiveFAROne}{86 }
\newcommand{\OFourATotalNumberPastroPointFiveFAROneBBH}{84 }

\begin{document}

\title{GWTC-4.0: Searches for Gravitational-Wave Lensing Signatures}

% FIXME commenting out for faster compilation
%% LVK authorlist in AAS format

%\collaboration{3000}{\LVKCollabAuthors}
%\affiliation{LIGO Hanford Observatory, Richland, WA 99352, USA}
%\affiliation{LIGO Livingston Observatory, Livingston, LA 70754, USA}
%\affiliation{Institute for Cosmic Ray Research, KAGRA Observatory, The University of Tokyo, 238 Higashi-Mozumi, Kamioka-cho, Hida City, Gifu 506-1205, Japan  }

\author[0000-0003-4786-2698]{A.~G.~Abac}
\affiliation{Max Planck Institute for Gravitational Physics (Albert Einstein Institute), D-14476 Potsdam, Germany}
\author{I.~Abouelfettouh}
\affiliation{LIGO Hanford Observatory, Richland, WA 99352, USA}
\author{F.~Acernese}
\affiliation{Dipartimento di Farmacia, Universit\`a di Salerno, I-84084 Fisciano, Salerno, Italy}
\affiliation{INFN, Sezione di Napoli, I-80126 Napoli, Italy}
\author[0000-0002-8648-0767]{K.~Ackley}
\affiliation{University of Warwick, Coventry CV4 7AL, United Kingdom}
\author[0000-0001-5525-6255]{C.~Adamcewicz}
\affiliation{OzGrav, School of Physics \& Astronomy, Monash University, Clayton 3800, Victoria, Australia}
\author[0009-0004-2101-5428]{S.~Adhicary}
\affiliation{The Pennsylvania State University, University Park, PA 16802, USA}
\author{D.~Adhikari}
\affiliation{Max Planck Institute for Gravitational Physics (Albert Einstein Institute), D-30167 Hannover, Germany}
\affiliation{Leibniz Universit\"{a}t Hannover, D-30167 Hannover, Germany}
\author[0000-0002-4559-8427]{N.~Adhikari}
\affiliation{University of Wisconsin-Milwaukee, Milwaukee, WI 53201, USA}
\author[0000-0002-5731-5076]{R.~X.~Adhikari}
\affiliation{LIGO Laboratory, California Institute of Technology, Pasadena, CA 91125, USA}
\author{V.~K.~Adkins}
\affiliation{Louisiana State University, Baton Rouge, LA 70803, USA}
% ...

%\author{The LIGO Scientific Collaboration}
%\author{The Virgo Collaboration}
%\author{The KAGRA Collaboration}

\date[\relax]{Compiled: \today}

\begin{abstract}
Gravitational waves can be gravitationally lensed by massive objects along their path. Depending on the lens mass and the lens--source geometry, this can lead to the observation of a single distorted signal or multiple repeated events with the same frequency evolution. We present the results for gravitational-wave lensing searches on the data from the first part of the fourth LIGO--Virgo--KAGRA observing run (O4a). We search for strongly lensed events in the newly acquired data by (1) searching for an overall phase shift present in an image formed at a saddle point of the lens potential, (2) looking for pairs of detected candidates with consistent frequency evolution, and (3) identifying sub-threshold counterpart candidates to the detected signals. Beyond strong lensing, we also look for lensing-induced distortions in all detected signals using an isolated point-mass model. We do not find evidence for strongly lensed gravitational-wave signals and use this result to constrain the rate of detectable strongly lensed events and the merger rate density of binary black holes at high redshift. In the search for single distorted lensed signals, we find one outlier: GW231123\_135430, for which we report more detailed investigations. While this event is interesting, the associated waveform uncertainties make its interpretation complicated, and future observations of the populations of binary black holes and of gravitational lenses will help determine the probability that this event could be lensed.

\end{abstract}


\section{Introduction}\label{sec:intro}

Massive astrophysical objects such as galaxies and galaxy clusters curve spacetime around them.
Propagating waves are deflected by such massive objects as they pass in their vicinity,
a phenomenon known as gravitational lensing.
According to the equivalence principle, gravitational lensing does not only affect light but also \ac{GWH} signals~\citep{Schneider:1992bmb}.
Strong gravitational lensing can produce repeated signals from the same source.
These repeated signals (referred to as images) can vary in amplitude, arrival time, and phase.
Due to the long wavelengths of \ac{GWH} signals detected by ground-based interferometers ($\sim 10^2$--$10^3$~km),
wave-optics effects can also be apparent when the characteristic size of the lens ($G M_\mathrm{L}/c^2$, with $G$ the gravitational constant and $c$ the speed of light) is comparable to the wavelength.
They may distort the waveform in characteristic ways~\citep{Takahashi:2003ix}.
The specific effects and intensity of gravitational lensing depend on the alignment of the source, lens,
and observer as well as the mass of the lens and the {GW} wavelength~\citep{Schneider:1992bmb,Takahashi:2003ix}. An introduction to {GW} lensing can also be found in Section 5.1.2. of \citet{LIGOScientific:IntroPaper}.

Identifying lensed {GW} signals presents numerous opportunities for probing new scientific avenues.
In cosmology, lensed {GW}s provide an independent method for measuring cosmological parameters~\citep{Liao:2017ioi,Hannuksela:2020tr,Finke:2021znb,Jana:2022shb, Balaudo:2022znx, Narola:2023viz,Wempe:2024jw,Jana:2024uta}, can break the mass-sheet degeneracy~\citep{Cremonese:2021puh, Chen:2024xal}, and
help probing large-scale structures~\citep{Mukherjee:2019wcg, Savastano:2023spl,Vujeva:2025kko} and the nature of dark matter~\citep{Tambalo:2022wlm,Jana:2024dhc,Jung:2017flg,Basak:2021ten,Urrutia:2021qak,barsode2024constraints}.
Additionally, repeated lensed {GW} signals allow for stringent tests of \acl{GR}~\citep[\acsu{GR};][]{Goyal:2020bkm,Ezquiaga:2020dao,Goyal:2023uvm,Chan:2025wgz}.
The improved sky localization of lensed GWs, especially when paired with electromagnetic counterparts,
enhances our ability to study their origin and environment~\citep{Hannuksela:2020tr,Uronen:2024bth}, and to provide early warning of mergers~\citep{Magare:2023hgs}, to give a few examples of the opportunities offered by multi-messenger lensing~\citep{Smith:2025axx}.

Continuous effort has been made to identify lensing signatures in {GW} data~\citep[e.g.,][]{Hannuksela:2019kle,Dai:2020tpj},
particularly during the \ac{O3} of the LIGO--Virgo--KAGRA (LVK) network~\citep{LIGOScientific:2021izm,LIGOScientific:2023bwz}.
Dedicated studies, including the follow-up lensing analyses of \ac{O3} events~\citep{Janquart:2023mvf}, have been conducted to identify
potential lensing candidates. However, no conclusive evidence for lensed GWs has yet been found.
With the improving sensitivity of the detectors~\citep{KAGRA:2013rdx, PhysRevD.111.062002, LIGO:2024kkz,LIGOScientific:IntroPaper}, the probability of detecting lensed GW signals is also
increasing as a typical relative strong lensing rate is $\mathcal{O}(1/1000)$~\citep{Ng:2017yiu,Li:2018prc, Mukherjee:2021qam, Wierda:2021upe,Xu:2021bfn}. In this work, we present searches for {GW} lensing
in the data collected during the \acl{O4a}~\citep[\acsu{O4a};][]{LIGOScientific:2025snk} and corresponding to new detections reported in GWTC-4.0~\citep{LIGOScientific:ResultsPaper}.

This paper is structured as follows.
In Section~\ref{sec:data}, we summarize the {GW} data collected during \ac{O4a},
and explain the selection of compact binary coalescence (CBC) events analyzed in this work.
Section~\ref{sec:framework_analysis} describes the analysis framework used to search for gravitational-lensing signatures,
including pairwise analyses for strong-lensing and single-event analyses for strongly-lensed type II images and wave-optics effects.
Section~\ref{sec:stronglensing} presents the results from pair-wise strong lensing searches, while
Section~\ref{sec:microlensing_and_millilensing} details the search for lensing in single signals. Section~\ref{sec:gw231123} reports on detailed investigations for GW231123\_135430 (hereafter referred to
as GW231123) identified as an outlier for our analyses in Section~\ref{sec:microlensing_and_millilensing} and~\citet{LIGOScientific:2025rsn}.
In Section~\ref{sec:implications}, we interpret the implications of our results in the context of astrophysical strong lensing rates and merger rate density.
Finally, Section~\ref{sec:conclusion} provides a summary of our findings and outlines prospects for future observing runs. The associated data release provide analysis results and figure generation scripts~\citep{LensingDataReleaseO4a}.

\section{Data and Events}\label{sec:data}

GWTC-4.0~\citep{LIGOScientific:ResultsPaper} is a cumulative catalog of {GW} detections to date, covering the first three observing runs and \ac{O4a}. It contains \GTWCFourTotalNumberPastroPointFive new \ac{O4a} candidates with a probability of astrophysical origin  \pastro $> 0.5$ and satisfactory event-validation checks~\citep{LIGOScientific:MethodsPaper}. \ac{O4a} ran from 2023 May 24 until 2024 January 16 with the two LIGO detectors~\citep{Capote:2024mqe, LIGO:2024kkz} taking data.
Of the \ac{O4a} candidates, \OFourATotalNumberPastroPointFiveFAROne pass a \ac{FAR} $ < 1\, \text{yr}^{-1}$ threshold. The candidates were identified in offline searches by different search pipelines detailed in~\citet{LIGOScientific:MethodsPaper}. The majority of these candidates are \acp{BBH}. In that period, 2 candidates, GW230529\_181500~\citep{LIGOScientific:2024elc} and GW230518\_125908, are consistent with sources containing a secondary mass $m_2 < 3 M_\odot$, making them likely \ac{NSBH} candidates. 
Candidates passing these BBH and FAR criteria have their source properties characterized using parameter estimation~\citep{LIGOScientific:MethodsPaper}. Here, we mostly rely on parameter-estimation results obtained using \textsc{IMRPhenomXPHM-SpinTaylor}~\citep{Pratten:2020ceb,Colleoni:2024knd} as base reference unlensed waveform, unless otherwise specified.

The analyses presented in this paper consider only candidates from \ac{O4a} with \ac{FAR} $ < 1\, \text{yr}^{-1}$ which are identified as \acp{BBH}, corresponding to a total of \OFourATotalNumberPastroPointFiveFAROneBBH new candidates. We do not include candidates from previous observing runs~\citep{Hannuksela:2019kle,LIGOScientific:2021izm,LIGOScientific:2023bwz} in our searches due to the computational cost, and because the time gap between the end of third observing run and the start of the fourth is too long to expect identifying strongly-lensed images spread between the two~\citep{Wierda:2021upe, Caliskan:2022wbh}, even if strong lensing by galaxy clusters could produce such time delays in some scenarios~\citep{Smith:2017mqu,Robertson:2020mfh,Ryczanowski:2020mnr}.
We do not include non-\ac{BBH} candidates as their lensing probability is low and GW230529\_181500~\citep{LIGOScientific:2024elc} has been previously investigated~\citep{10.1093/mnras/staf049}.

\section{Data-Analysis Framework}\label{sec:framework_analysis}

We use multiple analysis methods, each looking for specific lensing signatures.
The communication between each part of the analyses is automated in the \textsc{LensingFlow} framework~\citep{Wright:LFPaper}, which relies on
\textsc{Asimov}~\citep{Williams:2022pgn} and \textsc{CBCFlow}~\citep{CBCflow}. Parameter-estimation-based frameworks~\citep{Lo:2021nae,Janquart:2021nus,Janquart:2023osz,Wright:2021cbn} rely on \textsc{Bilby}~\citep{Ashton:2018jfp} with the \textsc{Dynesty} sampler~\citep{Speagle:2020wve}. For such analyses, both the \ac{PSD} and the priors on the source parameters were chosen to be consistent with the unlensed investigations carried out in~\citet{LIGOScientific:ResultsPaper}.  

As explained in Section 5.10 of~\citet{LIGOScientific:MethodsPaper}, in the late preparation of this work, a normalization error in the likelihood used for inference was found. Depending on the analysis requirements, different approaches (reweighting older posteriors or rerunning the corrected code) are used to mitigate this issue. This is detailed in the relevant sections. In the future, we will update the results affected by this error for all analyzed candidates affected by this issue.

\subsection{Searches for Multiple Images}\label{subsection:strong_lensing}

If the lens is massive enough, lensing effects can be described in the geometric-optics 
approximation. If the source is well-aligned with the lens, the GW is split into multiple distinct images with the same frequency evolution. Due to the limited angular resolution of GW detector networks, they will appear to be originating from the same region in the sky.
These images are delayed in time, (de)magnified, and can undergo 
an overall phase shift. So, the lensed waveform for image $j$ ($h_{\mathrm{L}}^{j}$) is linked to the unlensed waveform 
($h_{\mathrm{U}}$) as~\citep{Schneider:1992bmb} 
\begin{equation}\label{eq:strong_lensing}
 \tilde{h}_{\mathrm{L}}^j(f; \boldtheta, \boldphi_j) = \sqrt{\mu_j} ~ \tilde{h}_{\mathrm{U}}(f; \boldtheta) ~ e^{-2i\pi f t_j + i \pi n_j \mathrm{sign}(f) } \,,
\end{equation}
where $\boldtheta$ represents the usual \ac{BBH} parameters, and $\boldphi_j = \{\mu_j, t_j, n_j\}$ the lensing parameters,
with $\mu_j$ the magnification, $t_j$ the time delay, and $n_j$ the Morse factor~\citep{Schneider:1992bmb}. To get this relation, we follow the Fourier transform convention presented in Appendix B of~\citet{LIGOScientific:IntroPaper}. The Morse factor can only take one of the 
three following discrete values $\{0, 0.5, 1\}$, corresponding to an image forming at a minimum, a saddle point, or a maximum
of the lensing time delay, referred to as type I, II, and III images, respectively. This additional phase shift can lead 
to detectable features in {GW} signals, provided the image is of type II and the signal has signatures of precession, higher-order multipole moments, and/or eccentricity
~\citep{Dai:2017huk,Ezquiaga:2020gdt,Wang:2021kzt,Janquart:2021nus,Vijaykumar:2022dlp}.

We look for pairs of signals with matching characteristics. We do this for pairs of super-threshold candidates, i.e., new GWTC-4.0 \ac{BBH} candidates with $\mathrm{FAR} < 1 \, \mathrm{yr}^{-1}$ (Section~\ref{subsection:super_threshold_pairs}). Since lower magnification and changes in observing conditions can lead to images being below the detection threshold, we also look for counterpart sub-threshold images to the super-threshold candidates in the data (Section~\ref{subsubsection:sub_threshold}).

\subsubsection{Super-Threshold Image Pairs}\label{subsection:super_threshold_pairs}

When doing pair-wise image searches for strong lensing, one looks for pairs of events with matching time--frequency evolution. 
This means that parameters unaffected by lensing, such as the detector-frame masses or the spins, should be similar. 

Our workflow is split in multiple stages, referred to as tiers, with increasing computational cost and accuracy at each step. It starts with two tier-1 analyses, probing all the pairs made from the \ac{O4a} \ac{BBH} candidates, corresponding to a total of 3486 unique pairs. These are:
\begin{itemize}
    \item \textsc{\acl{PO}}~\citep[\textsc{PO};][]{Haris:2018vmn}, which looks at the consistency between the inferred properties of the two events for a subset of parameters (detector-frame component masses~\citep{LIGOScientific:MethodsPaper}, spin amplitudes and tilts, and the sky localization), as encoded in a ranking statistic $B_{\mathrm{U}}^{\mathrm{L}}$,
    combined with another ranking statistic $\mathcal{R}_{\mathrm{U}}^{\mathrm{L}}$ based on the expected time-delay distribution for galaxy lensing~\citep{Haris:2018vmn,More:2021kpb};
    \item \textsc{Phazap}~\citep{Ezquiaga:2023xfe}, which looks at the consistency of the reconstructed phases of the $\ell = m = 2$ mode of the GW signal at a specific frequency of 40 Hz, the arrival time between detectors, and the frequency evolution of the phases.
\end{itemize}

Each of these methods produces a specific ranking statistic that can be used to assess whether the event pair under consideration is sufficiently significant to be further analyzed; see Appendix~\ref{app:strong_lensing_ll_pipelines_and_result} for more detail. To make the comparison between different frameworks easier, we define a common \ac{FPP}\footnote{Similarly to~\citet{LIGOScientific:2023bwz}, we use FPP for significance associated with the lensing hypothesis (as opposed to the unlensed hypothesis), while FAR is associated with the significance assigned to individual candidate GW signal events (as opposed to them being noise events).} which can be computed based on an unlensed background distribution by taking the fraction of unlensed pairs with a ranking statistic higher than for the real pair under consideration. Information about the background used in this work can be found below, and more details are given in Appendix~\ref{app:background}. 

Pairs with a \ac{FPP} below 1\% are deemed interesting and passed to the next step of the workflow, the tier-2 analysis code \textsc{Fast-GOLUM}~\citep{Janquart:2021qov,Janquart:2023osz}. It evaluates the joint likelihood using samples obtained when analyzing the first image and linking them to the second image by sampling the relative magnification, time delay between the two images and the relative Morse factor difference. This recasting of \ac{JPE} is then faster than full \ac{JPE} in the next step since it is already focused on the region of interest, and
is further sped up by using a look-up table~\citep{Janquart:2021qov,Janquart:2023osz}. The \textsc{Fast-GOLUM} analysis is done using the \textsc{IMRPhenomXPHM-SpinTaylor} waveform~\citep{Pratten:2020ceb,Colleoni:2024knd} to which we add the lensing effects as described in Eq.~\eqref{eq:strong_lensing}, and a uniform prior between 0.1 and 50 for the relative magnification, a uniform prior between $[t_{ij}-0.2, t_{ij}+0.2]$~s for the time delay (where $t_{ij}$ is the difference in trigger times for the two events) and a discrete prior evenly distributed between the values $\{0, 0.5, 1, 1.5\}$ for the Morse factor difference. Priors on these parameters are chosen to cover a broad range of astrophysical lensing scenarios. Pairs are ranked using their coherence ratio $\mathcal{C}^\mathrm{L}_\mathrm{U} = p(d_1, d_2 | \mathrm{L})/\big(p(d_1|\mathrm{U}) p(d_2 | \mathrm{U})\big)$, where $d_i$ correspond to the data for the $i^{\rm th}$ image and ${\rm L}$ (${\rm U}$) refers to the lensed (unlensed) hypothesis. So, the coherence ratio corresponds to the ratio between the lensed and unlensed evidence for each {GW} signal~\citep{Lo:2021nae,Janquart:2021qov}. If $\log_{10}\mathcal{C}^\mathrm{L}_\mathrm{U} > 0$, the pair is passed to the next step.

For complete \ac{JPE}~\citep{Liu:2020par, Lo:2021nae,Janquart:2021qov,Janquart:2023osz}, corresponding to tier 3, we compute the Bayes factor using \textsc{Hanabi}~\citep{Lo:2021nae} and the \textsc{IMRPhenomXPHM-SpinTaylor} waveform~\citep{Pratten:2020ceb,Colleoni:2024knd}. It calculates the lensed evidence for the event pair by doing a joint analysis of the two data segments under the lensed hypothesis. The two signals are analyzed together with the same \ac{BBH} parameters and assuming they differ only through the lensing effect, with a change in amplitude, a time delay and an overall phase shift. \textsc{Hanabi} also reweighs the evidence to account for an astrophysical distribution of \ac{BBH} and lens parameters using the maximum-likelihood estimation of the GWCT-4.0 population~\citep{LIGOScientific:AstroDistPaper} and the same \ac{SIS} model as in~\cite{LIGOScientific:2023bwz}, respectively. It also includes selection effects~\citep{Lo:2021nae}. We refer to the final quantity including population models and selections effects as the strong lensing Bayes factor, following~\citet{Lo:2021nae} and \citet{LIGOScientific:2023bwz}.

\subsubsection{Sub-Threshold Image Search}\label{subsubsection:sub_threshold}

For a given super-threshold image, we can look for possible sub-threshold counterparts. 
To dig out such faint signals, one decreases the trials factor of the 
search by constructing a template bank containing a reduced number of waveforms. This is done by generating waveforms resembling the main signal~\citep{McIsaac:2019use,Li:2019osa,Dai:2020tpj,Li:2023zdl}. 

Sub-threshold searches are done with two approaches.
One relies on \textsc{PyCBC}~\citep{DalCanton:2014hxh,Usman:2015kfa,Nitz:2017svb}
to search for counterparts of each super-threshold event
with a single template~\citep{McIsaac:2019use},
which is chosen here as the aligned-spin projection of a maximum-posterior estimate from the \textsc{IMRPhenomXPHM-SpinTaylor} results.
Another method, \textsc{TESLA-X}, relies on \textsc{GstLAL}~\citep{Messick:2017lwm,Cannon:2021jlw} to do the search and 
construct a targeted template bank based on a dedicated injection campaign~\citep{Li:2023zdl, Li:2019osa}. 
\textsc{TESLA-X} results are only presented for GW231123, as detailed in Section~\ref{sec:gw231123}. 
Both methods matched-filter the HL-coincident times in the full O4a data set using the \textsc{SEOBNRv4} waveform model~\citep{Bohe:2016gbl}.
Further, to reduce the risk of spurious triggers, they veto triggers that do not overlap in 90\% credible sky localization (i.e., we require $\mathcal{O}_\text{90\% CI}>0$) with their targeted super-threshold candidate~\citep{Wong:2021lxf}. Pairs made of these triggers and their corresponding super-threshold event need to be followed up with the same methods as super-threshold pairs to assess the lensing hypothesis (Section~\ref{subsection:super_threshold_pairs}).

\subsection{Searches for Single Distorted Lensed Signals}\label{subsection:microlensing_and_millilensing}
We also report the results of analyses searching for single distorted signals
among the \ac{O4a} BBH candidates. One is a search for type II images produced by
strong lensing in the geometric-optics limit. The second is an analysis in the wave-optics regime using an isolated point mass lens model. The third is a targeted
phenomenological search for lensing near a fold caustic, which is applied only for specific candidates. In some cases, lensing-induced distortions can lead to the non-detection of signals with unlensed searches, leading to a selection effect~\citep{Chan:2024qmb}. This is not considered here. 

\subsubsection{Type II Image Searches}\label{subsubsection:type_II}

Type II image searches are done with the \textsc{GOLUM} framework~\citep{Janquart:2021qov,Janquart:2023osz} as in~\citet{LIGOScientific:2023bwz}. Even though such signals are expected to be rare in current detections~\citep{LIGOScientific:ResultsPaper}, we still search for such signatures as they could be unambiguous evidence for lensing~\citep{Ezquiaga:2020gdt,Wong:2021lxf,Janquart:2021nus,Vijaykumar:2022dlp} or mimic beyond-\ac{GR} effects~\citep{Mishra:2023vzo,Gupta:2024gun,Wright:2024mco,Narayan:2024rat}. Finding such a signal would imply the presence of other images. However, those may arrive at periods when detectors are offline or could be too faint to be detectable even by our sub-threshold searches.

\subsubsection{Searches for Wave-Optics Effects by Point Mass Lenses}\label{subsubsection:microlensing}

When the gravitational radius of the lens is comparable to the wavelength of the {GW} (wave-optics regime), lensing leads to frequency-dependent beating patterns in the signal. This can be modeled by an amplification factor $F(f; \boldphi)$ modifying the unlensed waveform~\citep{Takahashi:2003ix}
\begin{equation}\label{eq:microlensing}
 \tilde{h}_\mathrm{L}(f; \boldtheta, \boldphi) = F(f; \boldphi) ~  \tilde{h}_\mathrm{U}(f; \boldtheta) \,,
\end{equation}
where $\boldtheta$ represents the usual binary parameters, and $\boldphi$ the lensing parameters. The latter can be phenomenological or depend on the lensing model which is assumed. For the simplest mass profiles, like the isolated point-mass model used in this work~\citep{Schneider:1992bmb}, it corresponds to the redshifted lens mass $M_{\mathrm{L}}^z = (1+z_\mathrm{L})M_\mathrm{L}$, where $z_\mathrm{L}$ is the redshift of the lens and $M_\mathrm{L}$ its mass, and the impact parameter $y$. 
This model is chosen as it is expected to appropriately represent isolated compact lenses, ranging from isolated stars to black holes.
We compare the Bayes factor $\mathcal{B}^{\textrm{Mod}}_{\textrm{U}}$ with those obtained for an unlensed background to assess the related \ac{FPP} since imperfections in the data and models could lead to spurious favoring of the lensing hypothesis.

This analysis is done using the \textsc{Gravelamps} framework~\citep{Wright:2021cbn}, which incorporates lens models into \textsc{Bilby}~\citep{Ashton:2018jfp}. We use an updated framework compared to previous work~\citep{LIGOScientific:2021izm, LIGOScientific:2023bwz,Janquart:2023mvf}; an inconsistency in the Fourier transform convention was identified in the results obtained from past observing runs~\citep{Wright:NoteFix}. Updated results for past observing runs will be reported in separate work.

\subsubsection{Phenomenological Searches for Single Distorted Signals}\label{subsubsection:millilensing}

The transition from wave- to geometric-optics effects can be seen as a continuous change from one regime to another. Therefore, for isolated lenses with larger masses, there is a regime where the lensed {GW} signal can be modeled using a superposition of multiple images, each described by Eq.~\eqref{eq:strong_lensing} but where the time delay is at most a few tens of milliseconds. This leads to the observation of a single distorted signal, which can be modeled as a sum of all images, leading to a modification of the amplification factor in Eq.~\eqref{eq:microlensing} as~\citep{Liu:2023ikc}:
\begin{equation}\label{eq:millilensing}
F(f; \boldphi) = \sum_{j=1}^{K} \sqrt{\mu_j} ~ e^{-2i\pi f t_j + i \pi n_j \mathrm{sign}(f)} \,,
\end{equation}
where $\boldtheta$ represents the usual binary parameters, and $\boldphi = \{\boldphi_j\}_{j = 1, \dots, K}$ are the phenomenological lens parameters (magnification, time delay and Morse factor for each image), with $K$ the total number 
of images formed. 

A specific case of this kind of analysis, which is the one considered in this work, is that of lensing near a fold caustic, where one expects two lensed images ($K = 2$) with the same (large) magnification, opposite parities, and a millisecond time delay~\citep{Schneider:1992bmb, Lo:2024wqm,Ezquiaga:2025gkd}. 
We do not perform this analysis on all events, but consider it in more detail when investigating GW231123 (see Section~\ref{subsec:fold_caustic}). 

\subsection{Unlensed Background}\label{subsec:background}

We often compare the statistics obtained for the lensed versus unlensed hypothesis for each real event with that obtained for an unlensed background. This is done to compute \ac{FPP} values while accounting for fluctuations due to the noise present. Most of the lensing models have extra degrees of freedom compared to the unlensed one, potentially leading to a larger evidence for the lensed hypothesis even when the signal is not lensed but when additional features, like non-Gaussian noise fluctuations, are present. 

Our background is constructed using real stretches of \ac{O4a} data devoid of detected {GW} signals. We add an unlensed signal to the data and follow the usual analysis approaches~\citep{LIGOScientific:MethodsPaper}. 
We use 254 unlensed signals with masses sampled from the \textsc{PowerLaw + Peak} mass distribution, their spin magnitude from a Beta distribution, and their spin tilts from an isotropic plus truncated half-Gaussian mixture~\citep{KAGRA:2021duu}. All other parameters are drawn from their usual priors~\citep{LIGOScientific:MethodsPaper}. Injection and parameter estimation are done using the \textsc{IMRPhenomXPHM-SpinTaylor} waveform~\citep{Pratten:2020ceb,Colleoni:2024knd}, and parameter estimation follows the usual approaches~\citep{LIGOScientific:MethodsPaper}. 
Details about the background construction and analysis can be found in Appendix~\ref{app:background}.

\section{Searches for Multiple Images}\label{sec:stronglensing}

In this section, we report the results from searches for multiple images produced by strong lensing (Section~\ref{subsection:strong_lensing}).

\subsection{Super-Threshold Pairs}\label{subsec:sup_threshold_search} 

\begin{figure*}[ht]
    \centering 
    \includegraphics[keepaspectratio, width=0.8\textwidth]{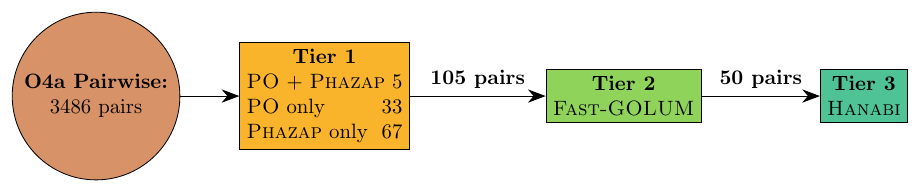}
    \caption{Flowchart showing the progression of the pairwise analyses and the numbers of pairs that passed each stage of the analysis process. With this approach, only a reduced number of pairs needs to be analyzed with JPE, significantly reducing the computational cost.}
    \label{fig:sl-flowchart}
\end{figure*}

The strong-lensing analyses are summarized in Figure~\ref{fig:sl-flowchart}, where the flowchart also reports the number of pairs analyzed and seen as interesting at the various steps in our workflow. These steps are further detailed below. Of the total 3486 initial pairs from all O4a events, only 50 were passed to our tier-3 analyses.

\subsubsection{Identification of Pairs with Tier-1 Analyses}\label{subsubsec:general_pairs_pairs}

Here, we report the results found with \textsc{PO}~\citep{Haris:2018vmn} and \textsc{Phazap}~\citep{Ezquiaga:2023xfe}, described in Section~\ref{subsection:super_threshold_pairs}. To account for the \textsc{Bilby} likelihood issue reported in Sec. 5.10 of~\citet{LIGOScientific:MethodsPaper}, we run these analyses on reweighted posteriors since changes are not expected to be large.

\begin{figure}
    \centering
    \includegraphics[width=\columnwidth]{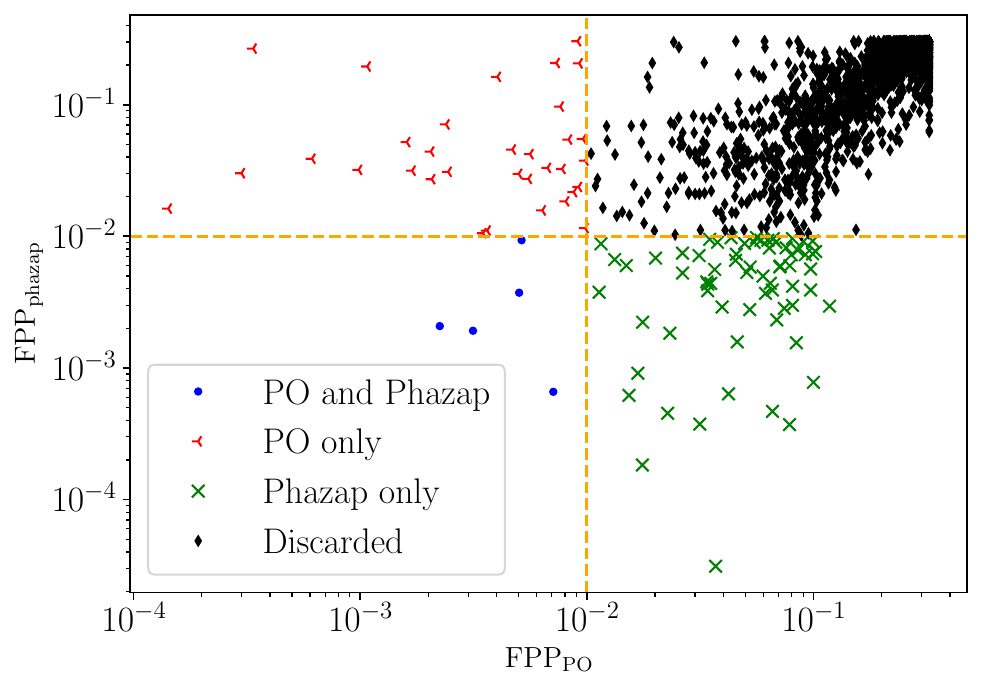}
    \caption{False-positive probability found for all O4a event pairs for \textsc{Posterior Overlap (PO)} and \textsc{Phazap}. The dashed orange line shows the 1\% threshold used to select pairs for follow-up. We pass to the second tier all pairs with an FPP below this threshold for at least one of the two approaches.}
    \label{fig:ll_pipelines_fpp}
\end{figure}

Figure~\ref{fig:ll_pipelines_fpp} shows the \ac{FPP} found for the two analyses run on all O4a events. 
Using a \textsc{FPP} threshold of 1\% for the two tier-1 approaches, we find a total of 105 pairs that need to be followed up by the tier-2 analysis, representing a 97\% reduction in the number of candidates. From all the pairs seen as interesting, there are 5 that are common to both analyses, 33 flagged only by \ac{PO} and 67 found only by \textsc{Phazap}. Details on the statistics used to select the pairs and compute their \ac{FPP}s can be found in Appendix~\ref{app:posterior_overlap} and \ref{app:phazap}. 

\subsubsection{Tier-2 Analysis}\label{subsubsec:mid_latency_analysis}

For this analysis, the runs are done using a likelihood fixed for the issue reported in Section 5.10 of~\citet{LIGOScientific:MethodsPaper}.
Out of the 105 pairs passed to \textsc{Fast-GOLUM}~\citep{Janquart:2021qov,Janquart:2023osz}, 55 pairs ($\sim 52\%$) were discarded, leading to 50 event pairs being passed to the tier-3 analyses. Details on the analysis and the results are given in Appendix~\ref{app:fastgolum}. 

\subsubsection{Tier-3 Analysis Results}\label{subsubsec:high_latency_results}

\begin{figure*}[h!]
\centering
\includegraphics[width=1.8\columnwidth]{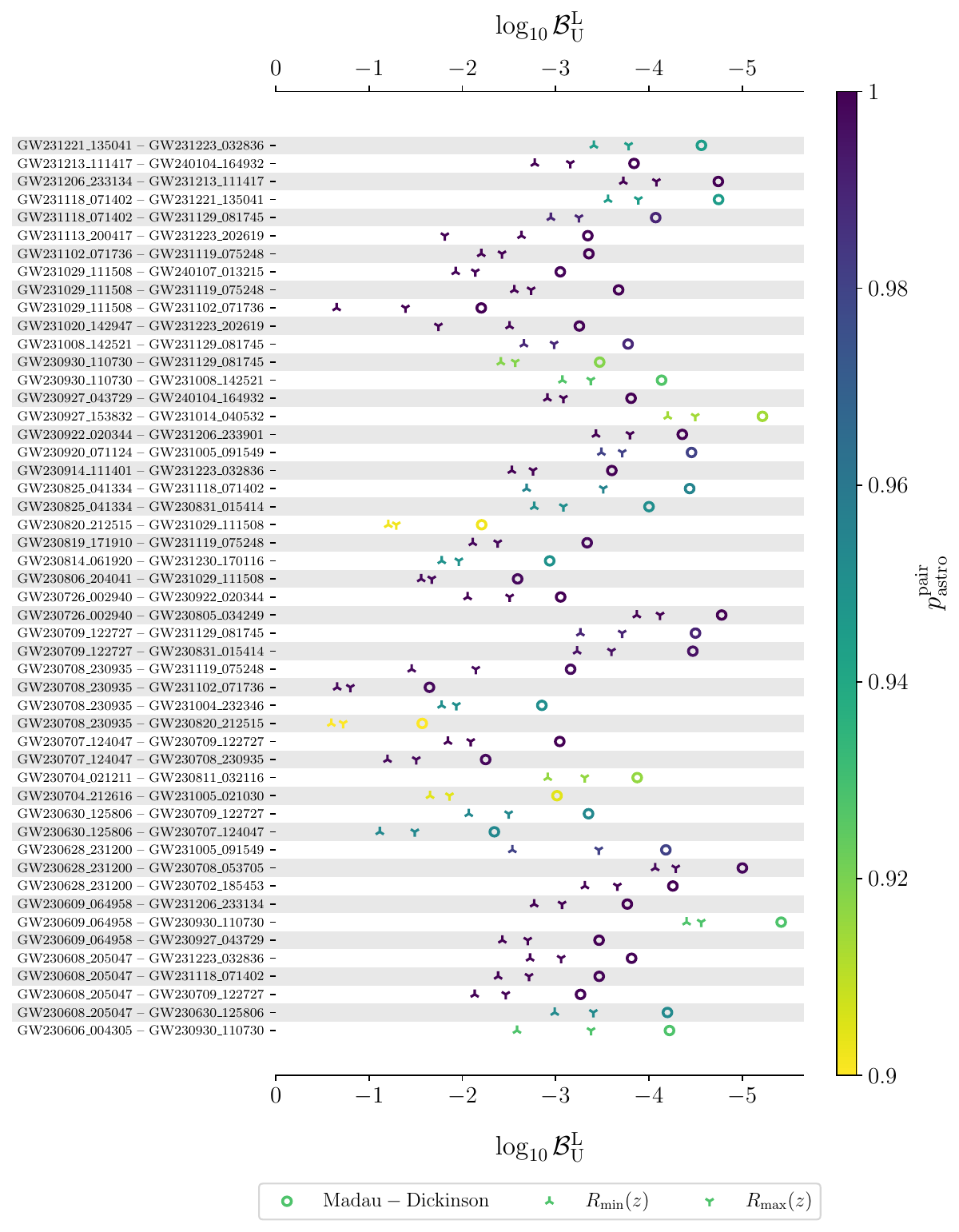}
\caption{\label{fig:strong_lensing_Bayes_factors}Bayes factors $\mathcal{B}^{\rm L}_{\rm U}$ comparing the lensed and unlensed hypotheses for the 50 event pairs passed to the \textsc{hanabi} analysis, computed using three different merger rate density models, namely one that tracks the star formation rate from~\citet{Madau:2014bja}, $R_{\rm min}(z)$ and $R_{\rm max}(z)$ that are described in~\citet{LIGOScientific:2023bwz}, respectively. The color of each marker indicates the value of $p_{\rm astro}^{\rm pair}$, which is the product of the probabilities that each event in a pair has an astrophysical origin as reported in~\citet{LIGOScientific:ResultsPaper}. None of the event pairs shows a preference for the strong-lensing hypothesis over the null hypothesis that the events in a pair are unrelated.}
\end{figure*}

For \textsc{Hanabi}, analyses are done using a likelihood fixed for the issue reported in Sec. 5.10 of~\citet{LIGOScientific:MethodsPaper}.
Out of the 3486 initial \ac{O4a} pairs, 50 were followed up by \textsc{Hanabi}~\citep{Lo:2021nae}.
\textsc{Hanabi} finds that none of the event pairs shows a preference for the strong-lensing hypothesis over the null hypothesis that the events in a pair are unrelated.

Since the strong-lensing Bayes factors are sensitive to assumptions about the population of the sources, we repeat the same Bayes factor calculation using three different models for the merger rate density, namely one that tracks the star formation rate from~\citet{Madau:2014bja}, as well as $R_{\rm min}(z)$ and $R_{\rm max}(z)$ that are described in~\citet{LIGOScientific:2023bwz}, respectively.
The corresponding markers in Figure~\ref{fig:strong_lensing_Bayes_factors} show the Bayes factors $\mathcal{B}^{\rm L}_{\rm U}$ computed using these merger rate density models.
We see that our conclusion that there is no evidence for strongly lensed {GW} signals within the 50 event pairs passed to the analysis is robust against differences in the assumed merger rate density.

The analysis assumes all of the GW events in those 50 pairs have astrophysical origins.
This might not be the case, and our assessment that an event is astrophysical in origin is encoded in $p_{\rm astro}$, reported in~\citet{LIGOScientific:ResultsPaper}.
To take this into account, we color each marker in Figure~\ref{fig:strong_lensing_Bayes_factors} with  $p_{\rm astro}^{\rm pair}$, which is the product of $p_{\rm astro}$ for each event in a pair.
All of the pairs we considered in the analysis have $p_{\rm astro}^{\rm pair} > 0.9$.

\subsection{Searches for Sub-Threshold Counterparts}\label{subsec:sub_threshold_search_analysis}

Searches for sub-threshold candidates potentially matching known GWTC-4.0 events were performed over all \ac{O4a} strain data
following the rules for data selection described in~\citet{LIGOScientific:MethodsPaper} and using the \textsc{PyCBC}-based targeted search algorithm \citep{McIsaac:2019use}.
For all sub-threshold searches, the templates are generated based on the posterior samples obtained prior to reweighing for the likelihood issue. This is not expected to change the results substantially; updated results are left as future work.

\begin{table*}
\centering
\begin{tabular}{l c r r c c r}
\hline
Sub-Threshold Trigger (UTC) & Target Super-Threshold Event & Time delay & Network SNR & FAR & $\mathcal{O}_\text{90\% CI}$ & $\mathcal{R}_{\mathrm{U}}^{\mathrm{L}}$\\
(yy:mm:dd hh:mm:ss) & GWYYMMDD\_HHMMSS & & & (yr$^{-1}$) & (\%) & \\
 \hline
 \hline
2023-09-03 07:43:35 & GW231104\_133418 & $62^\mathrm{d}05^\mathrm{h}50^\mathrm{m}44^\mathrm{s}$ & 7.88 & $5.6\times10^{-6}$ & 0.11 & 0.31 \\
2023-10-30 12:34:06 & GW231223\_075055 & $53^\mathrm{d}19^\mathrm{h}16^\mathrm{m}49^\mathrm{s}$ & 7.99 & $9.6\times10^{-6}$ & 52 & 0.41 \\
2023-11-20 02:21:03* & GW230904\_051013 & $76^\mathrm{d}21^\mathrm{h}10^\mathrm{m}50^\mathrm{s}$ & 9.62 & $2.5\times10^{-3}$ & 7.1 & 0.21 \\
2023-10-02 14:39:16* & GW230824\_033047 & $39^\mathrm{d}11^\mathrm{h}08^\mathrm{m}29^\mathrm{s}$ & 9.34 & 0.018 & 14 & 0.68 \\
2023-08-13 14:49:14* & GW231226\_101520 & $134^\mathrm{d}19^\mathrm{h}26^\mathrm{m}06^\mathrm{s}$ & 8.02 & 0.17 & 2.6 & 0.08 \\
2023-12-31 12:01:47* & GW230831\_015414 & $122^\mathrm{d}10^\mathrm{h}07^\mathrm{m}34^\mathrm{s}$ & 9.05 & 0.21 & 6 & 0.09 \\
2023-11-10 17:17:31* & GW231001\_140220 & $40^\mathrm{d}03^\mathrm{h}15^\mathrm{m}11^\mathrm{s}$ & 8.39 & 0.25 & 14 & 0.66 \\
2023-12-20 17:34:06* & GW230803\_033412 & $139^\mathrm{d}13^\mathrm{h}59^\mathrm{m}54^\mathrm{s}$ & 7.60 & 0.33 & 22 & 0.07 \\
2023-10-07 18:27:39 & GW231118\_071402 & $41^\mathrm{d}12^\mathrm{h}46^\mathrm{m}22^\mathrm{s}$ & 12.76 & 0.41 & 7 & 0.62 \\
2023-12-31 10:55:39* & GW230928\_215827 & $93^\mathrm{d}12^\mathrm{h}57^\mathrm{m}13^\mathrm{s}$ & 7.92 & 0.42 & 16 & 0.15 \\
2023-10-26 13:07:04* & GW230927\_043729 & $29^\mathrm{d}08^\mathrm{h}29^\mathrm{m}35^\mathrm{s}$ & 7.97 & 0.62 & 4.9 & 1.02 \\
2023-08-22 23:03:37* & GW230630\_125806 & $53^\mathrm{d}10^\mathrm{h}05^\mathrm{m}31^\mathrm{s}$ & 8.17 & 0.51 & 6 & 0.41 \\
2023-08-17 21:23:49* & GW230726\_002940 & $22^\mathrm{d}20^\mathrm{h}54^\mathrm{m}09^\mathrm{s}$ & 7.75 & 0.54 & 71 & 1.37 \\
2023-09-26 08:45:30 & GW231001\_140220 & $ 5^\mathrm{d}05^\mathrm{h}16^\mathrm{m}50^\mathrm{s}$ & 7.91 & 0.63 & 45 & 5.04 \\
2023-08-19 01:22:51* & GW231104\_133418 & $77^\mathrm{d}12^\mathrm{h}11^\mathrm{m}27^\mathrm{s}$ & 8.06 & 0.64 & 9.5 & 0.21 \\
2023-07-03 11:41:11* & GW230814\_061920 & $41^\mathrm{d}18^\mathrm{h}38^\mathrm{m}09^\mathrm{s}$ & 8.27 & 0.67 & 20 & 0.62 \\
2023-11-26 01:09:28* & GW231213\_111417 & $17^\mathrm{d}10^\mathrm{h}04^\mathrm{m}50^\mathrm{s}$ & 8.75 & 0.68 & 4.9 & 1.85 \\
2023-08-30 04:32:38* & GW231129\_081745 & $91^\mathrm{d}03^\mathrm{h}45^\mathrm{m}07^\mathrm{s}$ & 8.01 & 0.81 & 11 & 0.15 \\
2023-06-25 21:16:55* & GW231008\_142521 & $104^\mathrm{d}17^\mathrm{h}08^\mathrm{m}27^\mathrm{s}$ & 7.93 & 0.82 & 64 & 0.12 \\
2023-09-04 04:12:45* & GW240107\_013215 & $124^\mathrm{d}21^\mathrm{h}19^\mathrm{m}31^\mathrm{s}$ & 7.67 & 0.93 & 62 & 0.09 \\
2023-06-09 01:08:24* & GW230608\_205047 & $ 0^\mathrm{d}04^\mathrm{h}17^\mathrm{m}37^\mathrm{s}$ & 7.61 & 0.96 & 1.6 & 31.42 \\
2023-09-02 22:45:55* & GW231119\_075248 & $77^\mathrm{d}09^\mathrm{h}06^\mathrm{m}52^\mathrm{s}$ & 7.69 & 0.99 & 15 & 0.21 \\
 \hline
\end{tabular}
\caption{
The top 22 triggers
found by the \textsc{PyCBC} sub-threshold searches for the \OFourATotalNumberPastroPointFiveFAROneBBH
target super-threshold events,
in terms of lowest single-template \ac{FAR} below $1\,\text{yr}^{-1}$
and requiring sky overlap with the target event of $\mathcal{O}_\text{90\% CI}>0$.
The columns are: the date and time of the sub-threshold trigger,
the target event name from GWTC-4.0,
the time delay between the two,
the \ac{SNR} and \ac{FAR} of the new trigger in the single-template targeted search,
the sky localization overlap statistic $\mathcal{O}_\text{90\% CI}$,
and the time-delay ranking statistic $\mathcal{R}_{\mathrm{U}}^{\mathrm{L}}$.
Many of these triggers are found in multiple targeted searches.
For these,
we have only listed the lowest-\ac{FAR} result with non-zero sky overlap,
with the full set of triggers for all target events
included in the data release for this paper~\citep{LensingDataReleaseO4a}.
Triggers marked by an asterisk already appear in the full GWTC-4.0 candidate list,
though they do not pass the $\mathrm{FAR} < 1 \, \text{yr}^{-1}$ and $p_\mathrm{astro} > 0.5$ criteria.
}
\label{table:subthresh-search-results-top}
\end{table*}

In Table~\ref{table:subthresh-search-results-top}, 
we list all triggers from the \textsc{PyCBC} targeted searches for counterparts of the \OFourATotalNumberPastroPointFiveFAROneBBH
super-threshold targets that have a single-template 
\ac{FAR} below $1\,\text{yr}^{-1}$
and a non-zero sky overlap. We have also 
removed those corresponding to GWTC-4.0 events with
$\mathrm{FAR} < 1 \, \text{yr}^{-1}$ and $p_\mathrm{astro} > 0.5$.
Due to the high density of {GW} events in some regions of parameter space,
many triggers are recovered multiple times by searches for different super-threshold target events.
The table includes only the lowest-\ac{FAR} pairing for such cases.
Results of searches focusing on GW231123 are reported separately in Section~\ref{sec:gw231123},
and none of the \textsc{PyCBC} triggers for that event made this overall top list.
An extended list of all triggers from all searches below a \ac{FAR} of $12$\,yr$^{-1}$ is included in the data release for this paper~\citep{LensingDataReleaseO4a}.

Most of the triggers in Table~\ref{table:subthresh-search-results-top}
and 69 out of the 177 unique triggers in the data release
already appear in the full GWTC-4.0 candidate list,
though they correspond to triggers not satisfying the $\ac{FAR} < 1\,\text{yr}^{-1}$ 
and $p_\mathrm{astro}> 0.5$ criteria,
and generally do not have full data-quality and parameter-estimation information available.
On the other hand, the two lowest-\ac{FAR} triggers, found on
2023-09-03 07:43:35 (possible counterpart to GW231104\_133418)
and
2023-10-30 12:34:06 (possible counterpart to GW231223\_075055),
are not included in GWTC-4.0.
Both appear with low \ac{FAR} in only this targeted search.

As in previous studies~\citep[e.g.,][]{LIGOScientific:2023bwz}, we do not perform in-depth data-quality 
checks for possible instrumental origins of any of these triggers.
The reported \acp{FAR} also do not indicate how likely each trigger is 
an actual lensed counterpart of the target signal, but only how likely noise 
produces a trigger with a ranking statistic higher or equal to the one
under consideration using these single-template searches.
For our two top triggers, 
we ran parameter estimation with a setup similar to the one used 
for online candidates and assessed the probability of 
them being lensed counterparts to their corresponding target super-threshold event, using \textsc{PO} and \textsc{Phazap}.
We found that neither of
these two pairs match each other well enough to pass the threshold from Section~\ref{subsubsec:general_pairs_pairs},
and hence they are not interesting enough to be passed to further analyses.

More in-depth analyses could be carried out to select candidates of interest 
from this data release and study how much of the observed excess is due to noise artifacts, 
unlensed weak {GW} signals, or actual lensed counterparts.
To facilitate such studies~\citep{Ng:2024ooy}, we provide two additional statistics of interest for each super-threshold--sub-threshold pair:
the sky localization overlap statistic $\mathcal{O}_\text{90\% CI}$~\citep{Wong:2021lxf}
and the time-delay ranking statistic $\mathcal{R}_{\mathrm{U}}^{\mathrm{L}}$
based on galaxy-lensing priors~\citep{Haris:2018vmn, More:2021kpb}.

\section{Searches for Single Distorted Signals}\label{sec:microlensing_and_millilensing}

\subsection{Searches for Type II Images}\label{subsec:type_II_images}

\begin{figure}[t]
    \centering
    \includegraphics[width=0.5\textwidth]{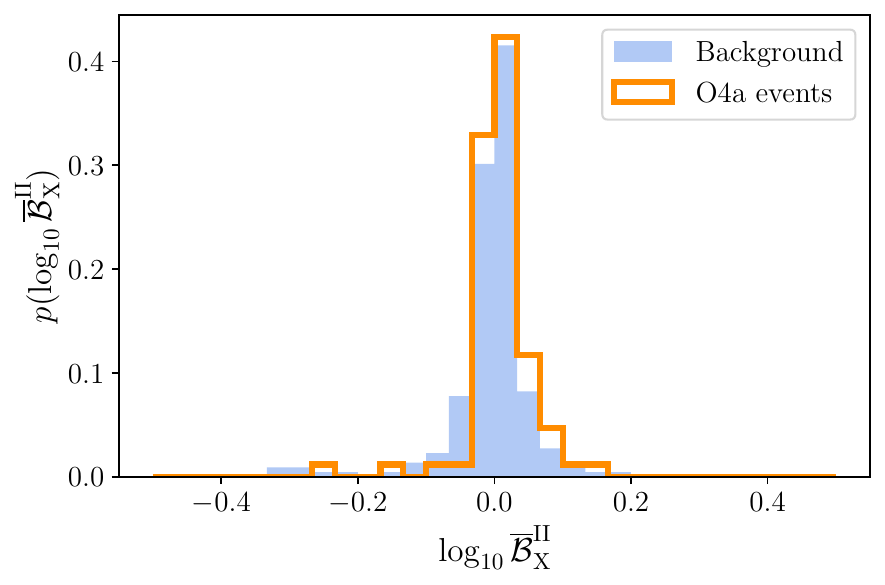}
    \caption{Distribution of averaged Bayes factors for the type II versus type I or III hypotheses obtained for real data and the unlensed background. No significant outlier candidates are identified.}
    \label{golum_typeII}
\end{figure}

Here, we report on the analyses looking for type II images detailed in Section~\ref{subsubsection:type_II}. Reported results are reweighted to account for the issue with the likelihood previously explained. Figure~\ref{golum_typeII} shows the distribution of averaged Bayes factors comparing the probability of a type II image as opposed to another image type, i.e., we take the mean value of the Bayes factor comparing type II vs type I images and that comparing type II vs type III images. This statistic is used because one cannot distinguish the difference between type I and III images, and any difference in Bayes factors is only due to sampling effects. Our results are overlaid with the distribution of Bayes factors from the unlensed background described in Section~\ref{subsec:background}. We find no outliers, meaning that no candidate shows support for the type II image hypothesis beyond what can be expected from noise fluctuations. 

\subsection{Point-Mass Lens Model Analyses}\label{subsec:microlensing} 

All O4a events were analyzed using the isolated point-mass model for the amplification~\citep{Wright:2021cbn}, and the \textsc{IMRPhenomXPHM-SpinTaylor} waveform model~\citep{Pratten:2020ceb,Colleoni:2024knd}.
We use a uniform prior between $1\,M_\odot$ and $10^{4}\,M_\odot$ for the redshifted lens mass ($M^{z}_{\rm L}$), while the prior for the dimensionless source position ($y$) is $\propto y$ with $y \in [0.1, 3]$.
The mass range corresponds to lens masses from stars ($\sim 1M_{\odot}$) to intermediate-mass black holes ($\sim 10^{4}M_{\odot}$), 
beyond which the time delay between images becomes large enough to resolve multiple signals. 

To handle the likelihood issue mentioned in Sec.~\ref{sec:framework_analysis}, for all \textsc{Gravelamps} analyses and complementary unlensed analyses, results were obtained by reweighing the nested sampling chains rather than the posterior distributions. This leads to more robust calculation of the evidence.

\begin{figure}[t]
    \centering
    \includegraphics[width=1.0\linewidth]{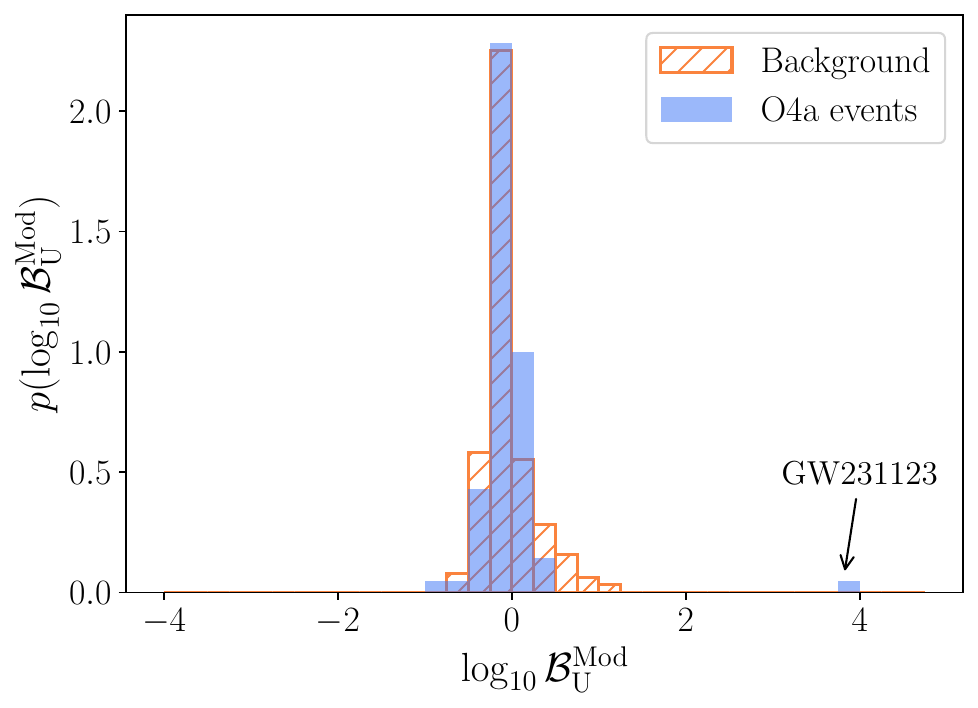}
    \caption{Histogram of $\textrm{log}_{10}\mathcal{B}^{\textrm{Mod}}_{\textrm{U}}$ for O4a BBH candidates. Except for GW231123, all Bayes factors found with IMRPhenomXPHM-SpinTaylor fall within the range expected for unlensed events based on the background.}
    \label{hist_micro_bf}
\end{figure}

For each event, we calculate a Bayes factor $\mathcal{B}^{\textrm{Mod}}_{\textrm{U}}$, corresponding to the evidence ratio between the two hypotheses. The resulting distribution is shown in Figure~\ref{hist_micro_bf}.
To assess statistical fluctuations in the Bayes factor for unlensed signals, we also show the values of $\mathcal{B}^{\textrm{Mod}}_{\textrm{U}}$ for the astrophysical background described in Section~\ref{subsec:background} and Appendix~\ref{app:background}.
The distribution of Bayes factors for O4a events is comparable to that of the background except for GW231123, 
which has $\rm{log}_{10}\mathcal{B}^{\textrm{Mod}}_{\textrm{U}}=3.8$. This is the highest value measured so far~\citep{LIGOScientific:2025rsn} and will be analyzed in more detail in Section~\ref{sec:gw231123}. Moreover, since this value is outside of our astrophysical background, we can place an upper bound on its FPP of $0.39\%$, where the statistic is defined as the ratio between unlensed signals with a higher Bayes factor to all unlensed events in the background and does not include the trials factor. The $p$-value including trials factor is given by 
\begin{equation}
 p = 1 - (1 - \textrm{FPP})^{N} \, ,
\end{equation}
where $N$ is the number of events. For our case, we find $p < 28\%$, showing the limitations related to the size of our astrophysical background. Another potential limitation of our background is the low abundance of GW231123-like events, for which there is a higher risk of false alarm due to degeneracies between spin and lensing effects~\citep{liu2024can}. Additionally, short signals are known to be more affected by non-Gaussian noise fluctuations~\citep[e.g.,][]{LIGOScientific:2025cmm, Ray:2025rtt}, which could then be captured by the lensing model. While our background uses real noise to account for such effects, the noise was cleaned before injections were done (see Section~\ref{app:background}), which could lead to over-cleaning and reduce the effect of non-Gaussian features. So, further studies are needed to quantify their impact. 

Because of its outstanding properties, we investigate GW231123 in more detail in the next section.

\section{GW231123\_135430}\label{sec:gw231123}

This section presents more thorough analyses of GW231123, flagged in
\citet{LIGOScientific:2025rsn} as a potential candidate for lensing distortions, which is also further
shown in Section~\ref{sec:microlensing_and_millilensing}. We focus on comparing the
lensed and unlensed hypotheses for a quasi-circular \ac{BBH} coalescence.

Under the unlensed hypothesis, GW231123 is exceptional as its source
corresponds to one of the most massive \acp{BBH} detected so far, with a
total mass of $190$--$265 \, M_\odot$ and high
spins~\citep{LIGOScientific:2025rsn}. Analysis results also show variation with the waveform model. By standard accuracy measures, \textsc{NRSur7dq4} is, on average, the most accurate available model in this region of parameter space, although~\citet{LIGOScientific:2025rsn} finds that in some cases all models can lead to biased parameter estimates.

Here, we present the properties of the source and point mass lens under the lensed hypothesis, evaluate the evidence of lensing by fold caustic, search for various strong lensing effects, discuss the astrophysical priors of lensing, investigate in more detail waveform consistency in the lensing context, and look at possible noise artifacts that could increase the support for the lensing hypothesis. 

\subsection{Properties of GW231123 and the Corresponding Point Mass Lens} \label{subsec:gw231123_lensing_properties}

For the single-distorted signal analyses done, GW231123 has the highest Bayes
factor seen so far, above what is expected from our astrophysical background.
Noting the significant disagreements between the source parameters estimated 
using different (unlensed) waveform approximants~\citep{LIGOScientific:2025rsn}, we similarly perform our analyses using multiple
waveforms. This initial set of tests is performed using the
\textsc{IMRPhenomXPHM-SpinTaylor}~\citep{Pratten:2020ceb, Colleoni:2024knd},
\textsc{IMRPhenomXO4a}~\citep{Thompson:2023ase}, and
\textsc{NRSur7dq4}~\citep{Varma:2019csw} waveform models. The corresponding recovered Bayes factors are shown in Table~\ref{tab:Bayes_factors_231123}. 
Note that we consider only an 
isolated point mass lens and do not consider the case of e.g., a microlens embedded
in the potential of a bigger lens, such as a galaxy, which could change the
inferred quantities~\citep{Diego:2019lcd, Cheung:2020okf, Yeung:2021chy,
Mishra:2021xzz,
Mishra:2023ddt,Shan:2022xfx,Shan:2023qvd,Shan:2023ngi,Shan:2024min,seo2022improving, seo2025residual}.
While these Bayes factors indicate some favoring of the lensing hypothesis
under the assumption of Gaussian and stationary noise for the
\textsc{IMRPhenomXPHM-SpinTaylor} and \textsc{NRSur7dq4} cases, support is more
marginal for \textsc{IMRPhenomXO4a}.
Spurious favoring of lensing can be caused by
features unaccounted in the analyses, such as non-Gaussian noise effects or inaccuracy in waveforms used. 
The favoring of lensing decreases for waveform models more favored by the data~\citep{LIGOScientific:2025rsn}. This shows a possible support for lensing coming from compensating for differences between the waveform models and the data. We follow up on some of these aspects in this and the next subsections.

\begin{table}
\centering
    \begin{tabular}{lccc}
        \hline
        Waveform Model & XPHM-ST & NRSur & XO4a\\
        \hline
        \hline
        $\log_{10}(B_\mathrm{U}^{\mathrm{Mod}})$  & 3.8 & 1.5 & 0.1 \\ 
        \hline
    \end{tabular}
    \caption{Bayes factors for GW231123 when comparing the point-mass lensed and unlensed models with three different waveform
    approximants.
    XPHM-ST, NRSur, and XO4a denote \textsc{IMRPhenomXPHM-SpinTaylor}, \textsc{NRSur7dq4}, and \textsc{IMRPhenomXO4a}, respectively.}
    \label{tab:Bayes_factors_231123}
\end{table}

For the point-mass analysis, we infer the lens properties (the redshifted lens mass and the dimensionless impact parameter) using the three waveform models. Using \textsc{IMRPhenomXPHM-SpinTaylor} (\textsc{NRSur7dq4}), the redshifted lens mass and dimensionless impact parameter are $911^{+504}_{-292}M_{\odot}$ ($803^{+489}_{-299}M_{\odot}$) and $0.61^{+0.27}_{-0.21}$ ($0.69^{+0.37}_{-0.26}$), respectively. The \textsc{IMRPhenomXO4a} waveform yields broader posterior ranges for both lens parameters, with $687^{+13208}_{-356}M_{\odot}$ and $2.13^{+0.79}_{-1.54}$. Details of the lens properties are provided in Appendix~\ref{app:GW231123_pointmass_posteriors}.

\begin{figure}
  \centering
  \includegraphics[keepaspectratio, width=\columnwidth]{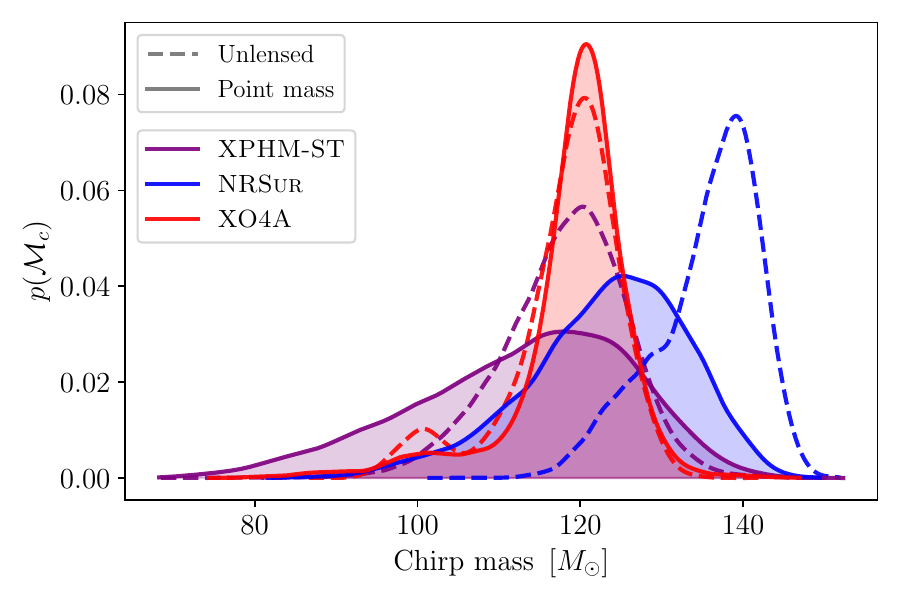}
  \caption{
    Comparison of the detector-frame chirp mass posteriors under the
    point mass lensing (solid line) and unlensing (dashed line)
    hypotheses using the three considered waveform models: \textsc{IMRPhenomXPHM} (\textsc{XPHM-ST}), \textsc{NRSur7dq4} (\textsc{NRSur}), and \textsc{IMRPhenomXO4a} (\textsc{XO4a}). The difference
    between the waveforms is reduced when the analysis is done under the lensing hypothesis. 
  }
  \label{fig:chirp_mass_example}
\end{figure}

Turning to the inferred BBH properties, under the lensed hypothesis, we observe a broadening of the posteriors and a reduction in the differences between the waveform models considered. In
Figure~\ref{fig:chirp_mass_example}, we show that apparent differences in the
recovered posterior in the unlensed analyses are reduced in the lensed ones
for the detector-frame chirp mass, with similar trends found for all parameters.
Broader posteriors are expected under the lensing hypothesis, because of the greater freedom in fitting the lens. The broader posteriors extend to regions of parameter space where waveform models are more reliable.

Analyses were conducted examining only the data obtained by each of the LIGO
Hanford (LHO) and LIGO Livingston (LLO) detectors. Results for each detector are then
compared to ascertain whether any transient noise artifacts in a single
detector could result in the apparent support for lensing, as was noted for
GW200208\_130117 in~\cite{Janquart:2023mvf}. 
Results from both detectors indicate support for the lensing hypothesis,
with stronger evidence in the LLO analysis than in the LHO analysis.
Specifically, the $\log_{10}(B_\mathrm{U}^{\mathrm{Mod}})$ values are 0.62 for LHO and 2.64 for LLO analyses with the \textsc{IMRPhenomXPHM-SpinTaylor} model. Such a difference could be driven by the higher SNR present in LLO compared to LHO, or to specific features present in one detector (see Section~\ref{subsec:noise_GW231123} for a discussion on this topic).
Furthermore, we calculate Kolmogorov–Smirnov (KS) statistics and Jensen–Shannon (JS) divergence between the chirp mass and spin amplitude posteriors obtained
from the unlensed LHO and LLO analyses and those from the corresponding lensed analyses.
As shown in Table~\ref{tab:single_ifo_231123}, both measures take lower values in the lensing case, showing the analyses under the lensed hypothesis show greater consistency for information obtained from each detector individually.

\begin{table}
\centering
\begin{tabular}{lcc cc cc}
    \hline
    Parameter & \multicolumn{2}{c}{$\mathcal{M}$} & \multicolumn{2}{c}{$\chi_{1}$} & \multicolumn{2}{c}{$\chi_{2}$} \\
           & KS & JSD & KS & JSD & KS & JSD \\
    \hline\hline
    Unlensed & 0.647 & 0.294 & 0.150 & 0.044 & 0.051 & 0.002 \\
    Lensed   & 0.342 & 0.089 & 0.097 & 0.009 & 0.026 & $4\times 10^{-4}$ \\
    \hline
\end{tabular}
\caption{Kolmogorov–Smirnov statistics (KS) and Jensen–Shannon divergence (JSD, in nats) values between the chirp mass and spin magnitude posteriors obtained for individual detectors under the unlensed and lensed hypotheses using the \textsc{IMRPhenomXPHM-SpinTaylor} waveform.}
\label{tab:single_ifo_231123}
\end{table}

\begin{figure}
    \centering
    \includegraphics[width=\linewidth]{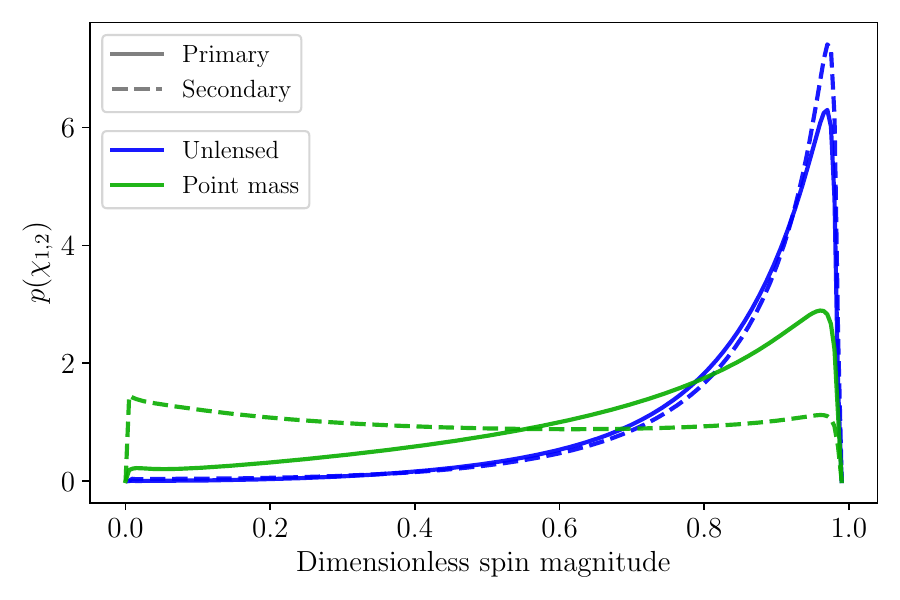}
	  \caption{
      Posteriors of the dimensionless spin magnitudes recovered for GW231123 in
      the lensed (green) and unlensed (blue) analyses with the \textsc{NRSur7dq4} waveform model. The solid(dashed) curves correspond to the
      primary (secondary) component.
      Under the lensing hypothesis, there is greater support for more moderate
      spins, particularly in the secondary component.
    }
    \label{fig:spin_magnitudes_combined}
\end{figure}

In the unlensed analysis, there is a distinct preference for extremal spins
for this event. Lensing distortions and the effects of spin precession are highly correlated~\cite[e.g.,][]{liu2024can}. The
comparison between the spins for the primary and secondary are shown in
Figure~\ref{fig:spin_magnitudes_combined} for the \textsc{NRSur7dq4} waveform model. The observed behavior is qualitatively similar for all waveform models considered here. Under the lensing hypothesis, the
higher values for the primary spin remain preferred albeit with a broader peak.
The posterior on the secondary becomes much broader, granting more support for
moderate spin values at the expense of support for extremal values. This
greater support for moderate spin amplitudes may also lead to the greater
agreement between the waveforms as it would bring the waveform models 
toward regions of better calibration~\citep{Varma:2019csw,Pratten:2020ceb,Colleoni:2024knd,Thompson:2023ase}.

Detector-frame masses remain high under the lensed hypotheses. 
When examining the source-frame masses under the isolated point mass model, these
would be $109^{+15}_{-13} M_{\odot}$ for the primary and $59^{+13}_{-11}
M_{\odot}$ for the secondary in the \textsc{NRSur7dq4} analysis. This
represents a modest drop in the primary mass and a significant decrease for the
secondary compared with the $129^{+15}_{-14} M_{\odot}$ and $114^{+14}_{-17}
M_{\odot}$ for the primary and secondary respectively reported from the
unlensed analysis~\citep{LIGOScientific:2025rsn}.

\subsection{Searching for Lensing Near a Fold Caustic}
\label{subsec:fold_caustic}

Here, we analyze GW231123 in the context of lensing near a fold caustic of a foreground lens, producing highly magnified images. 
This analysis has only one extra free parameter compared to the unlensed BBH case, which is the time delay $t_{ij}$ between the two lensed signals.
A uniform distribution from $0.002$ s to $0.6$ s is chosen to be the agnostic prior for $t_{ij}$. The lower bound is limited by the sampling rate of the data that we used, while the upper bound is chosen such that the entire lensed waveform will still fit inside the standard $8$ s analysis window.

\begin{table}
  \centering
\begin{tabular}{lc}
  \hline
	Waveform Model & $\log_{10}\mathcal{B}^{\rm{Fold}}_{\rm{U}}$ \\
	\hline
  \hline
	\textsc{IMRPhenomXPHM-SpinTaylor} & $\;\;\;2.05$ \\
	\textsc{NRSur7dq4} & $-0.05$ \\
  \hline
\end{tabular}
\caption{\label{tab:logBF_for_fold}Log Bayes factor for the lensing near a fold caustic scenario compared to the unlensed BBH merger for GW231123, using two different models for the underlying BBH waveform.}
\end{table}

Table \ref{tab:logBF_for_fold} shows the Bayes factor comparing the lensing near a fold caustic with the unlensed BBH merger using the same two underlying waveform models, \textsc{IMRPhenomXPHM-SpinTaylor}~\citep{Pratten:2020ceb, Colleoni:2024knd} and \textsc{NRSur7dq4}~\citep{Varma:2019csw}. 
We find that the unlensed hypothesis is preferred over the fold-caustic lensing when using \textsc{NRSur7dq4}, while the opposite is true when using \textsc{IMRPhenomXPHM-SpinTaylor}.
This is in line with other unlensed analyses showing discrepancies between the two models~\citep{LIGOScientific:2025rsn}.  

When comparing the recovered posterior distribution for the \ac{BBH} parameters, we also see that the two waveform models have a better agreement under the fold-caustic lensing hypothesis than under the unlensed hypothesis, possibly partly fitting information missing under the unlensed hypothesis. However, the obtained Bayes factors are not significant enough to claim strong evidence supporting fold-caustic lensing.

\subsection{Searching for Strong Lensing Effects}\label{subsec:strong-lensing}

Here we search for signatures of strong lensing in GW231123. We evaluate the evidence of it being a Type II lensed image, an outlier in the population of BBHs, and search for any sub-threshold counterparts produced by strong lensing. 

\subsubsection{Type II Lensed Image}\label{subsubsec:type-ii-lensed-image}

While no evidence for type II images were found in Section~\ref{subsec:type_II_images},
here, we re-do the \textsc{GOLUM} type II analyses for GW231123 with adapted waveform models and using two different priors on the Morse factor: a discrete one ($n_j \in \{0, 0.5, 1\}$) and a continuous one (between 0 and 1). The latter is done because if a signal is of type II, the posterior should peak at 0.5 while other effects, like noise, waveform systematics and deviation from GR, can lead to a Morse factor value peaking away, which cannot be seen in the discrete case~\citep{LIGOScientific:2023bwz,Janquart:2023mvf, Wright:2024mco}. We also do analyses on individual detectors to gauge the consistency in results between them. In addition to the Bayes factor comparing explicitly the image types ($\mathcal{\overline{B}}^{II}_{X}$), we also report the Bayes factor comparing the evidence obtained in the lensed and unlensed analyses ($\mathcal{B}^{\mathrm{L}}_{\mathrm{U}}$). The results are summarized in Table~\ref{table:golum_typeII}.

\begin{table}
\centering
\begin{tabular}{lcccc}
  \hline
  Waveform & Detectors & $n_{j}$ prior & $\log_{10}\mathcal{B}^{\mathrm{L}}_{\mathrm{U}}$ & $\log_{10}\mathcal{\overline{B}}^{II}_{X}$ \\
  \hline
  \hline 
  \textsc{NRSur} & LHO--LLO & discrete  & $0.04$ & $-0.80$ \\ 
\textsc{NRSur} & LHO--LLO & continuous  & $-0.45$ &  \\ 
\textsc{XPHM-ST} & LHO--LLO & discrete  & $-0.17$ & $-0.35$ \\ 
\textsc{XPHM-ST} & LHO-only & continuous  & $0.27$ &  \\ 
\textsc{XPHM-ST} & LHO--LLO & continuous  & $-0.14$ &  \\ 
\textsc{XPHM-ST} & LHO-only & discrete  & $0.22$ & $-0.01$ \\ 
\textsc{XPHM-ST} & LLO-only & continuous  & $0.29$ &  \\ 
\textsc{XPHM-ST} & LLO-only & discrete  & $-0.52$ & $0.51$ \\
\hline
\end{tabular}
 \caption{Summary table of the GOLUM type II analysis results for the \textsc{IMRPhenomXPHM-SpinTaylor} (\textsc{XPHM-ST}) and \textsc{NRSur7dq4} (\textsc{NRSur}) waveforms. The third row corresponds to the result for GW231123 plotted in Fig.~\ref{golum_typeII}. There are discrepancies between results depending on the waveform and detector configuration chosen. However, overall, no consistent evidence for a type II image is seen.}
    \label{table:golum_typeII}
\end{table}

Generally, while the \textsc{NRSur7dq4} waveform model disfavors type II images, the \textsc{IMRPhenomXPHM-SpinTaylor} results show greater ambiguity with support for type II images found in LLO-only runs, as well as the combined-detector continuous prior run. However, both the LHO--LLO and LLO-only continuous prior runs do not peak distinctly around $n_j = 0.5$, with some skew towards a higher or lower phase shift, thus indicating that this support is likely to come from some other, unidentified effect rather than true type II effects~\citep{Janquart:2023mvf,Wright:2024mco,Narayan:2024rat}. Type II analyses with different waveforms recover different posterior distributions for intrinsic parameters, supporting the systematics identified in~\citet{LIGOScientific:2025rsn}. In summary, we do not find consistent evidence for a type II image.

\subsubsection{Searches for Population Outlier}\label{subsubsec:pop_outlier}

If a strongly lensed image is not identified as lensed, its magnification will bias the inferred source properties~\citep{Oguri:2018muv}. In particular, a magnified event ($\mu>1$) will appear as closer and more massive. Therefore, it is possible to search for strongly lensed signals as potential outliers of the inferred astrophysical population~\citep{Hannuksela:2019kle, Farah:2025ews}. To avoid false positives, it is necessary to account for the uncertainty in the inferred astrophysical model as well as event parameter uncertainty induced by noise in the detector~\citep{Fishbach:2019ckx}. If inferred source properties are identified as outliers, the magnification needed to make it consistent with the population can be determined. Alternatively, one can assume the lensing hypothesis and then check if the inferred magnification is consistent with the strong-lensing likelihoods~\citep{Hannuksela:2019kle, LIGOScientific:2021izm}. However, large magnifications are needed to lead to observable outliers, and such occurrences are rare~\citep{Farah:2025ews}.

While GW231123 has exceptionally high mass, it is still consistent with the high-mass tail for the population distribution and is not an outlier~\citep{LIGOScientific:2025rsn,LIGOScientific:AstroDistPaper}. This fact alone is not sufficient to gauge the possibility that the event is lensed~\citep{Farah:2025ews}. 

\subsubsection{Sub-Threshold Counterparts}~\label{subsubsec:sub-threshold-counterparts}

If GW231123 is part of a multiplet of images, it is of interest to see if we can find sufficiently similar counterparts in the data. No super-threshold candidate pairs pass the selection criteria from Section~\ref{subsec:sup_threshold_search},
but it is possible that some sub-threshold counterparts exist (see Section~\ref{subsec:sub_threshold_search_analysis}). Here we report also lower-significance triggers for completeness. We report the results for these searches with the \textsc{TESLA-X}~\citep{Li:2019osa,Li:2023zdl} and \textsc{PyCBC}-based~\citep{McIsaac:2019use} searches.
Including triggers with \ac{FAR} below 1 per 30\,days,
the first approach leads to two triggers being found,
while the other leads to eight triggers that correspond to known super-threshold events
and four additional sub-threshold triggers,
only two of which have sky localization overlap with GW231123.
None of these triggers are common between the two searches.
The two \textsc{TESLA-X} triggers and one from \textsc{PyCBC} correspond to candidates included in the full GWTC-4.0 list but not as sufficiently significant for full data-quality and parameter-estimation studies.
The triggers with non-zero sky overlap are reported in Table~\ref{tab:gw231123-sub-threshold-candidates} with the same information as in Table~\ref{table:subthresh-search-results-top}.
None of them appear to be of sufficient interest based on their \acp{FAR} and we, therefore, do not perform further follow-up studies on them.

\begin{table*}[t!]
\centering
\begin{tabular}{l c c c c c c}
\hline
Analysis & Sub-Threshold Trigger (UTC) &  Absolute Time Delay & Network \ac{SNR} & \ac{FAR} & $\mathcal{O}_\text{90\% CI}$ & $\mathcal{R}_{\mathrm{U}}^{\mathrm{L}}$ \\
& (yy:mm:dd hh:mm:ss) & (days) & & (yr$^{-1}$) & (\%) \\
 \hline 
 \hline 
 \textsc{PyCBC} & 2023-10-29 05:30:40* & 25.35 & 8.15 & 4.13 & 16 & 1.22 \\
 \textsc{PyCBC} & 2023-07-30 17:22:24 & 115.86 & 8.61 & 4.87 & 11 & 0.10 \\
 \textsc{TESLA-X} & 2023-11-13 06:20:50* & 10.32 & 8.64 & 6.52 & 23 & 3.10 \\
 \textsc{TESLA-X} & 2023-10-02 14:39:16* & 51.97 & 9.41 & 9.97 & 19 & 0.43 \\
 \hline
\end{tabular}
\caption{Possible counterpart triggers found for GW231123 with our sub-threshold searches and with non-zero sky overlap, excluding those already considered as significant GWTC-4.0 events,
down to \ac{FAR} of 1 per 30\,days.
We find two such triggers with the \textsc{TESLA-X} search and four with the \textsc{PyCBC}-based search,
though two of those do not have overlapping sky localizations with GW231123 and are not reported in the table.
Triggers marked with an asterisk correspond to candidates
included in the full GWTC-4.0 list,
but due to $\ac{FAR} > 1\,\text{yr}^{-1}$
in the global search no data-quality or parameter-estimation studies were performed on those.
\textsc{PyCBC} also recovered eight of the significant GWTC-4.0 events
as possible counterparts for GW231123,
but those were already excluded by the standard
super-threshold lensing analyses discussed in this paper.
}
\label{tab:gw231123-sub-threshold-candidates}
\end{table*}

\subsection{Possible Astrophysical Systems}\label{subsec:prior_odds}

To assess the significance using Bayesian approaches, one should ideally evaluate the Bayes factor against the plausibility of lensing. Here, we consider a few selected example systems where lensing may be expected, based on existing literature.

From an astrophysical perspective, for a black hole lens, one would expect such systems to be formed in dense environments, such as star clusters for a central massive lens, von Zeipel--Lidov--Kozai induced mergers~\citep{Ubach:2025dob}, or mergers happening in a migration trap around an active galactic nucleus~\citep{Leong:2024nnx}. Such models require a particular formation channel for the BBHs. However, under those scenarios, \citet{Ubach:2025dob} and \citet{Leong:2024nnx} find that a small fraction of events may be lensed, down to $\mathcal{O}(10^{-4})$. Alternatively, one can take a more empirical approach and consider lensing candidates with millisecond to second time delays reportedly observed via gamma ray bursts~\citep{Paynter:2021wmb, Levan:2025ool}. While these candidates are being debated~\citep{Mukherjee:2023qkg}, they can offer an avenue to constraining the plausibility of the lensing hypothesis empirically. 

A comprehensive assessment of the astrophysical priors for lensing, including selection effects and considering specific lens models outlined above, should be carried out in the future to better assess the plausibility of the lensing hypothesis.

\subsection{Waveform Reconstruction}\label{subsec:cwb_comparison}

\begin{figure*}
  \centering
  \includegraphics[width=0.85\linewidth]{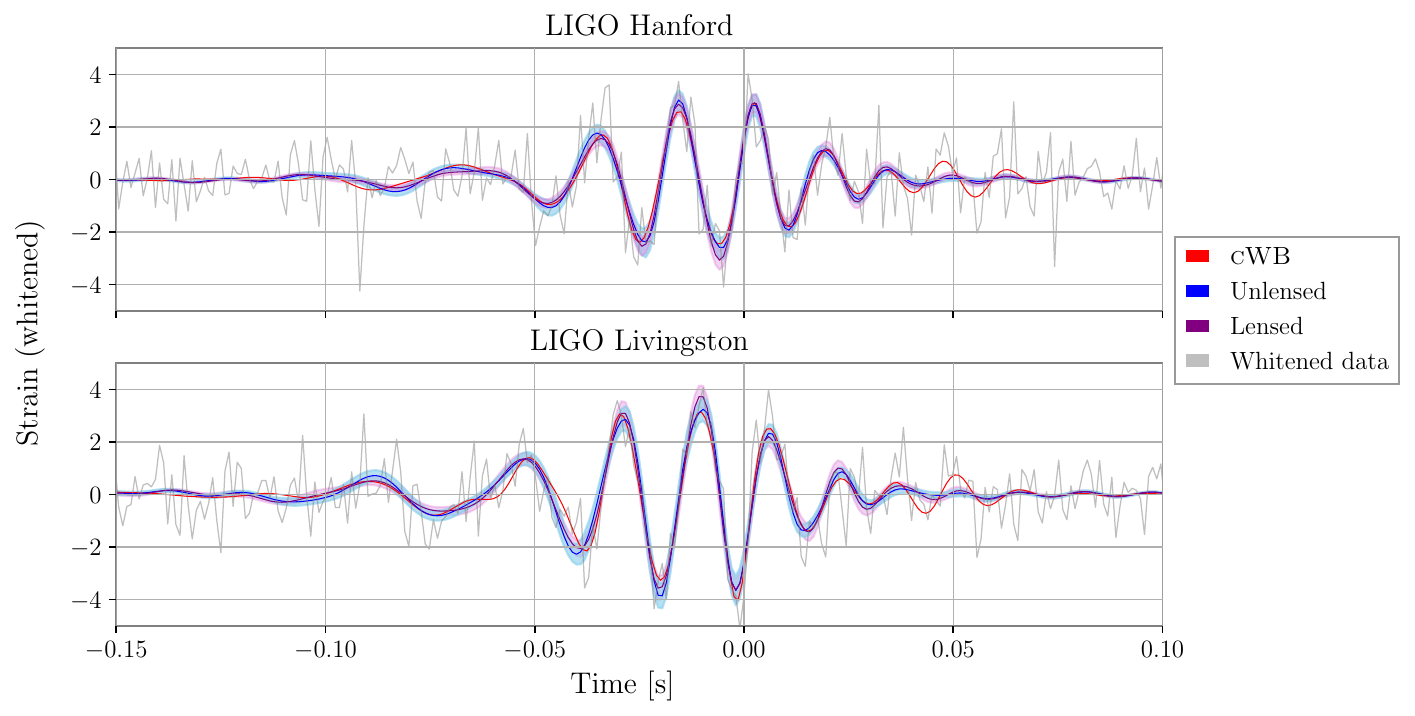}
  \caption{
    Comparison between the \textsc{cWB} waveform reconstruction and the 90\%
    credible interval waveforms from the parameter-estimation analyses under the
    lensed and unlensed hypotheses with the \textsc{NRSur7dq4} waveform model. The time represents the time relative to 2023 November 23 at 13:54:30.619 UTC. The lensed signal does not show an overall visibly closer match to the \textsc{cWB} waveform.
  }
  \label{fig:cwb-comparison}
\end{figure*}

One investigation to understand where the apparent support for the lensing
hypothesis (under the point mass lens model presented in Section~\ref{subsec:gw231123_lensing_properties}) could come from is to compare the inferred unlensed and lensed
waveforms with the one reconstructed by an unmodelled search pipeline for
GW signals~\citep{Shan:2023ngi}. Here, we use
\textsc{cWB}~\citep{Drago:2020kic}, which uses a wavelet basis to reconstruct
coherent power across detectors. A comparison of the waveforms from the
\textsc{cWB} analysis, the unlensed parameter estimation, and the
lensed parameter estimation (using the point mass lens model) are shown in
Figure~\ref{fig:cwb-comparison}. Here and in the next Section, we focus on \textsc{NRSur7dq4} since it was shown in~\citet{LIGOScientific:2025rsn} to be the most accurate waveform model on average for events in this region of the parameter space.

To quantify the difference between the two models, we use the mismatch, which represents the dissimilarity between two signals and is defined as~\citep{Owen:1995tm}
\begin{equation}
  \bar{\mathcal{M}} = 1 - \max_{t_0, \phi_0} \frac{\langle h_1 | h_2 \rangle}{\sqrt{\langle h_1 | h_1 \rangle \langle h_2 | h_2 \rangle}}\,,
\end{equation}
where $\langle h_1 | h_2 \rangle = \int \mathrm{d}f\, \tilde{h}^{*}_1(f)\tilde{h}_2(f)/S_n(f)$ is the noise-weighted inner product between two waveforms $h_1$ and $h_2$, with $\tilde{h}(f)$ being the Fourier transform of $h(t)$, $S_n(f)$ the one-sided power spectral density of the detector noise, and the maximization is done over time and phase of coalescence.
Looking at the mismatch between the \textsc{cWB} reconstruction and the recovered lensed and unlensed waveforms when analyzing the signal with \textsc{NRSur7dq4}, we find the lensed waveform to generally better match the model-agnostic reconstructed signal.
A distribution of mismatch values is given in Appendix~\ref{app:GW231123_waveform}, and we find mismatches of $0.045^{+0.010}_{-0.007}$ and $0.051^{+0.010}_{-0.009}$ for LHO and LLO for the unlensed model, while they are of $0.035^{+0.007}_{-0.005}$ and $0.039^{+0.010}_{-0.007}$ for the lensed case. The difference between the two being small, they are likely to be caused by the extra degrees of freedom in the lensed model and are probably not significant.

\subsection{Noise Considerations}\label{subsec:noise_GW231123}

A possible cause for the favoring of lensing would be the presence of non-Gaussian noise features.
There are known glitches, corresponding to noise over-densities, around the time of the signal. Therefore, the LHO analyses are done using deglitched data~\citep{LIGOScientific:2025rsn} but non-Gaussian noise features could still be present in the data biasing results, especially if considering population of microglitches~\citep{Ray:2025rtt}. 
In this case, one would need the noise fluctuations to be similar in the two detectors to give the observed consistency with lensing characteristics.

\begin{figure}
    \centering
    \includegraphics[height = 7cm, width=\columnwidth]{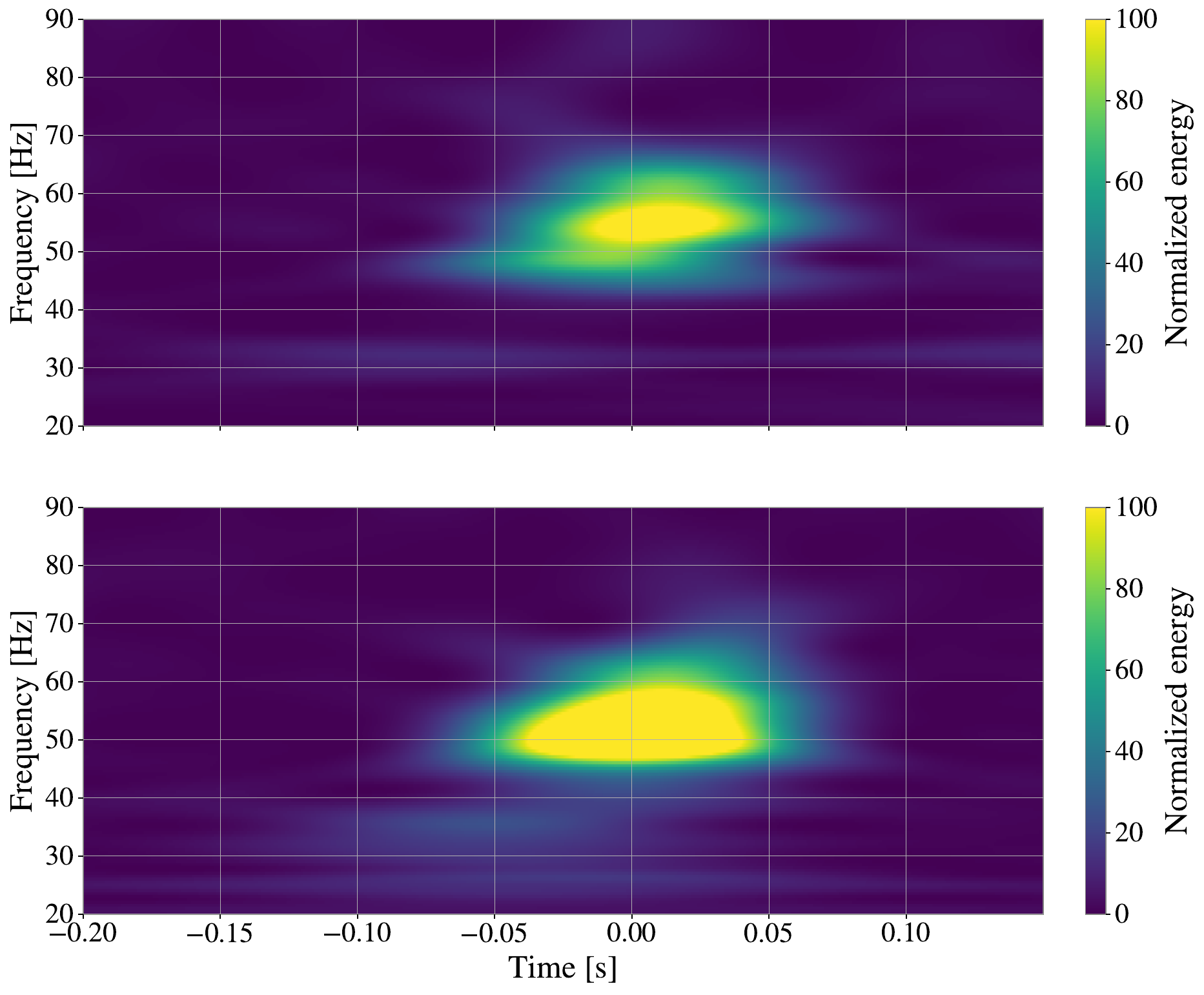}
    \caption{Normalized squared amplitude in time–frequency space for the lensed (top) and unlensed (bottom) results obtained with the NRSur7dq4 waveform model and the maximum-likelihood points. 
    The time represents the offset relative to 2023 November 23 at 13:54:30.619 UTC.
    A comparison between the two waveforms reveals differences spread across multiple frequency bins rather than confined to a single bin, and these do not coincide with a particular spectral line.}
    \label{fig:power_lensed_unlensed}
\end{figure}

While this is unlikely for random noise fluctuations, a possible candidate would be the 60 Hz line related to the power grid.
In that case, one would expect the lensing support to be significantly reduced if the line is removed and the main excess power from lensing
to be concentrated around that frequency. Doing such an analysis, we find that notching does not change the lensing support, which is also in line with findings from~\citet{LIGOScientific:2025rsn}, where results were unchanged after notching the 60Hz line.
Additionally, we verify whether the power is spread  and not concentrated in a narrow band around 60 Hz by examining the power difference between lensed and unlensed results.
Figure~\ref{fig:power_lensed_unlensed} shows the normalized squared amplitude in time-frequency space for the two cases at the maximum-likelihood points of the NRSur7dq4 analyses.
As can be seen, the lensed case exhibits a broader power distribution compared to the unlensed case, indicating that the extra fitting introduced by lensing affects multiple frequency regions simultaneously rather than being confined to a single narrow band.
Therefore, it is unlikely the support is due to a spectral line. 

\section{Implications}\label{sec:implications}
In this section, we present implications of the non-detection 
results of our strong-lensing search analyses, from which we can derive constraints on the detection rate of strong lensing and on the \ac{BBH} merger rate density at high redshift.

\begin{table*}[t!]
\begin{center}
\begin{tabular}{lcccc}
\hline
		& \multicolumn{2}{c} {Galaxy lenses}  & \multicolumn{2}{c}  {Galaxy Cluster lenses}\\ 
Merger Rate Density Model &  $R_{\rm D}$ & $R_{\rm S}$ & $R_{\rm D}$ & $R_{\rm S}$ \\ 
\hline\hline
\textsc{Power-Law } &  18.1$\times 10^{-4}$ & 24.5$\times 10^{-4}$ & 5.4$\times 10^{-4}$ & 7.6$\times 10^{-4}$ \\
\textsc{Power-Law + Peak} & 3.2--9.9 $\times~10^{-4}$   &   5.0--13.5 $\times~10^{-4}$   &   0.9--3.8 $\times~10^{-4}$   &   2.1--5.1 $\times~10^{-4}$ \\
\hline
\end{tabular}
\end{center}
\caption{
 \label{tab:lensedrates0_min}
 Expected fractional rates of observable lensed single or double events at current LVK sensitivity. Relative rates for lensed double events where $R_{\rm D}$ implies both of the lensed events are detected and $R_{\rm S}$ implies only the louder of the two events is detected above the \ac{SNR} threshold and computed for galaxy- and cluster-scale lenses separately. In the \textsc{Power-Law } model, the model parameters are set to the median of the posteriors constrained by GWTC-4.0~\citep{LIGOScientific:AstroDistPaper}. In the \textsc{Power-Law + Peak} model, the parameters are obtained from the combined analysis of CBC detections and limits from SGWB~\citep{LIGOScientific:IsotropicSGWB}. Here, the ranges in the lensing rates encompass a 90\% credible interval given the posteriors.}
\end{table*}

\subsection{Strong-Lensing Event Rates}\label{subsec:strong_lensing_rates}

We constrain the relative occurrence of lensed events due to galaxies and galaxy clusters compared to unlensed ones by following standard techniques~\citep{Ng:2017yiu,Li:2018prc, Mukherjee:2021qam, Xu:2021bfn,Wierda:2021upe,Phurailatpam:2024enk}. We simulate a large population of lensed and unlensed mergers based on the latest population constraints and determine the relative observation of such signals. The lens-mass model is chosen to be \ac{SIS} for all types of lenses. For the galaxies, the velocity dispersion function follows from the Sloan Digital Sky Survey~\citep{Choi:2006qg} with velocity dispersion $\sigma_{\rm vd}=10$--$350 \,{\rm km~s}^{-1}$. For galaxy clusters, we use the halo-mass function from~\citet{Tinker:2008ff} with halo masses $\log_{10}(M_{\rm halo}/M_\odot) \sim 14$--$16$. Although it could impact rates, we do not consider the effect of sub-substructures in clusters~\citep{Vujeva:2025kko}. The \ac{BBH} population is taken from~\citet{LIGOScientific:AstroDistPaper}. 
We choose a detection criterion on the two-detector SNR of the lensed events to be above 8. Furthermore, this \ac{SNR} cut is applied to (a) the weaker of the two lensed signals implying both are detectable (denoted D), or (b) the stronger of the two lensed signals implying at least one lensed signal is detectable (denoted S).  

Table~\ref{tab:lensedrates0_min} gives the latest median and $90\%$ strong-lensing event rates relative to the unlensed event detection rate for two choices of merger rate density models: (a) \textsc{Power-Law} model where the local rate $R_0$ and the power-law index $\kappa$ are set to the median of the posteriors that are constrained by GWTC-4.0~\citep{LIGOScientific:AstroDistPaper}, and (b) \textsc{Power-Law + Peak} model where the posteriors of the parameters are jointly constrained from the GWTC-4.0 results as well as the non-detection of the stochastic GW background~\citep[SGWB;][]{LIGOScientific:IsotropicSGWB}. In this case, the lensing rates are shown for the $90\%$ credible interval. Our results are consistent with those based upon GWTC-3.0~\citep{LIGOScientific:2023bwz} and have decreased uncertainties due to improved constraints on the merger rate density.

\subsection{Constraints on the Binary Black Hole Merger Rate Density at High Redshift}\label{subsec:bbh_merger_rate_density}

The lensing optical depth is larger for binaries at high redshifts; thus the non-detection of strongly lensed {GW} signals provides a unique way of constraining the \ac{BBH} merger rate at redshifts larger than the redshift horizon of the current detectors under the assumption that rate density models can be extrapolated to higher redshifts. This is complementary to the constraints obtained from the non-detection of SGWB, since the relative contribution of binaries at different redshifts to the SGWB and lensing is, in general, different~\citep{Buscicchio:2020cij, Buscicchio:2020bdq, Mukherjee:2020tvr}.

\begin{figure}
	\centering
	\includegraphics[keepaspectratio, width=0.5\textwidth]{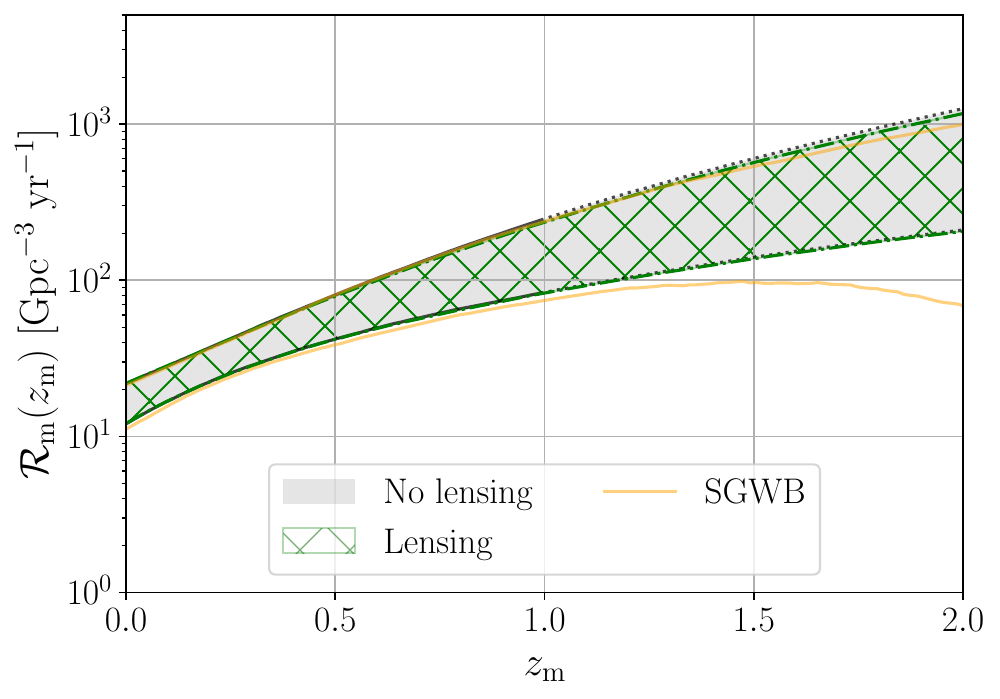}
	\caption{Constraints on the \ac{BBH} merger rate density as a function of the merger redshift ($z_\mathrm{m}$), obtained from the non-detection of strongly lensed \ac{GW} signals during O4a (green cross-hatched region). For reference, we show the currently allowed merger rates constrained by the GWTC-4.0 results~\citep[grey shaded region, no lensing;][]{LIGOScientific:AstroDistPaper}. The solid orange curves enclose the constraints from the absence of detection of a SGWB in the O4a data~\citep{LIGOScientific:IsotropicSGWB}.}
	 \label{fig:high_z_merger_rate}
\end{figure}

We start with the latest posteriors for the merger rate density model, assumed to have power-law redshift dependence and constrained by the GWTC-4.0~\citep{LIGOScientific:AstroDistPaper}, to retain those models that are consistent with non-detection of strongly lensed \ac{BBH} signals (see \cite{LIGOScientific:2021izm,LIGOScientific:2023bwz} for details). Figure~\ref{fig:high_z_merger_rate} shows the original GWTC-4.0 only constraints and the revised upper bounds from absence of strongly lensed \ac{BBH} events. Independent of the strong lensing analysis, we show constraints on the merger rate density obtained from the non detection of a SGWB for reference. The non-detection of lensing signatures provides upper limits comparable to those obtained from the SGWB. The SGWB results assume a \textsc{Power-Law + Peak} model for the redshift-dependence of the merger rate densities (as described in Section~\ref{subsec:strong_lensing_rates}) and hence, a direct comparison between strong lensing and SGWB is not trivial. More stringent constraints from strong lensing are expected in the future as detectors become more sensitive and more signals are detected~\citep{Buscicchio:2020cij,Buscicchio:2020bdq,Mukherjee:2020tvr}. 

\section{Conclusion}\label{sec:conclusion}
We have performed searches for GW lensing signatures in O4a data. 
Our analysis combined three complementary strands:

\paragraph{Pair-wise strong-lensing searches:} All $3486$ possible pairs of O4a astrophysical BBH detections with $\textsc{FAR} < 1\,\mathrm{yr}^{-1}$ were 
processed with two independent tier-1 analyses (\textsc{Posterior Overlap} and \textsc{Phazap}) 
and an intermediate fast-joint-PE filtering process (tier 2). After filtering, $50$ pairs remained and were followed up 
with a full joint parameter estimation (tier 3). All pairs considered have Bayes factors consistent with unlensed expectations, and none were further followed up.

\paragraph{Sub-threshold counterpart searches:} Targeted searches for possible sub-threshold lensed counterparts were conducted for all
BBH candidates found in O4a with $\textsc{FAR} < 1\,\mathrm{yr}^{-1}$. The general search was done with a \textsc{PyCBC}-based approach.
We have presented the candidates recovered with single-template FAR $< 1 \mathrm{yr}^{-1}$. We followed up on the new interesting candidates, not finding significant support for lensing. We also reported candidate counterparts for GW231123 found with two different approaches (\textsc{TESLA-X} and \textsc{PyCBC}).

\paragraph{Single-event analyses:} All BBH candidates with $\textsc{FAR} <
1\,\mathrm{yr}^{-1}$ were re-analysed with an isolated point-mass lens model, searching for frequency-dependent lensing distortions. 
With the exception of GW231123, no signal showed a significant lensing signature, and Bayes factors were undistinguishable from those of unlensed simulations.
No signal displayed the unambiguous type-II phase structure predicted for strongly lensed images with large higher-mode content.

\paragraph{GW231123:} This event was further investigated in order to understand the significance of the lensing hypothesis and to understand what could be driving the large Bayes factors. 
The probability of an astrophysical background of unlensed signals
producing this level of consistency with a lensed waveform is $< 0.39\%$ ($< 28\%$) without (with) the inclusion of the trials factor, where the significance is limited by the size of our background simulations. However, missing physics in the unlensed waveform models compensated by lensing effects could drive the large Bayes factors. Further observations of \acp{BBH} and of gravitational lenses will help determine the probability for GW231123 to be lensed.

\paragraph{Implication of the non-detection of strong lensing:} Assuming GW231123 is not affected by strong lensing, the non-detection of this lensing effect with BBH-population and lens-population models yields an 
observable lensed doubles per unlensed BBH detection rate of $3.2$--$9.9\times 10^{-4}$ ($0.9$--$3.8\times 10^{-4}$) for galaxy (galaxy-cluster) lenses at O4a sensitivity. Thus, the non-detection of strong lensing in the current data is not surprising. 
The absence of detected strongly lensed signals also enables constraints on the BBH merger rate density at a redshift larger than one. The upper limit found is comparable to that found via the non-detection of the SGWB. 

Future runs with higher sensitivity and more detectors~\citep{KAGRA:2013rdx} will also help in finding more confident detections of GW lensing signatures by reducing the risk of false matches with better sky localization and better measurement accuracy.

\section*{Data Availability}

All strain data analysed as part of GWTC-4.0 are publicly available through GWOSC~\citep{OpenData}. The details of this data release and information about the digital version of the GWTC are described in detail in~\citet{LIGOScientific:2025snk}. Data products associated with this work, including analysis results and plotting scripts, are available in \citet{LensingDataReleaseO4a}.

\section*{Acknowledgments}
Analyses in this catalog relied upon the \textsc{LALSuite} software
library~\citep{lalsuite}.
Estimates of the noise spectra and glitch models were obtained using
\textsc{BayesWave}~\citep{Cornish:2014kda,Littenberg:2015kpb,Cornish:2020dwh}.
The detection of sub-threshold candidates and their statistical significance were 
obtained using \textsc{GstLAL}~\citep{Messick:2016aqy,Sachdev:2019vvd,Hanna:2019ezx,Cannon:2020qnf} and \textsc{PyCBC}~\citep{DalCanton:2014hxh, Usman:2015kfa,Nitz:2017svb,Davies:2020tsx} based software.
Source-parameter estimation was performed
with the \textsc{Bilby}
library~\citep{Ashton:2018jfp,Smith:2019ucc,Romero-Shaw:2020owr} using the
\textsc{Dynesty} nested sampling package~\citep{Speagle:2020wve}.
\textsc{PESummary} was used to postprocess and collate parameter-estimation
results~\citep{Hoy:2020vys}.
The various stages of the parameter-estimation
analysis were managed with the \textsc{Asimov} library~\citep{Williams:2022pgn}.
For lensing analyses, we used \textsc{TESLA-X}~\citep{TESLA-X_git}, \textsc{Posterior Overlap}~\citep{PO_git}, \textsc{Phazap}~\citep{phazap_git}, \textsc{GOLUM}~\citep{GOLUM_git}, and \textsc{Gravelamps}~\citep{Gravelamps_git}.
Some of the software above rely on standard libraries such as \textsc{Astropy}~\citep{astropy:2022}, \textsc{Cython}~\citep{cython:2011}, and \textsc{Pandas}~\citep{mckinney-proc-scipy-2010}.
Plots were prepared with \textsc{Matplotlib}~\citep{Hunter:2007ouj} and
\textsc{Seaborn}~\citep{Waskom:2021psk}.
\textsc{NumPy}~\citep{Harris:2020xlr} and \textsc{SciPy}~\citep{Virtanen:2019joe} were used
in the preparation of the manuscript.

This material is based upon work supported by NSF’s LIGO Laboratory which is a major facility
fully funded by the National Science Foundation.
The authors also gratefully acknowledge the support of
the Science and Technology Facilities Council (STFC) of the
United Kingdom, the Max-Planck-Society (MPS), and the State of
Niedersachsen/Germany for support of the construction of Advanced LIGO 
and construction and operation of the GEO\,600 detector. 
Additional support for Advanced LIGO was provided by the Australian Research Council.
The authors gratefully acknowledge the Italian Istituto Nazionale di Fisica Nucleare (INFN),  
the French Centre National de la Recherche Scientifique (CNRS) and
the Netherlands Organization for Scientific Research, 
for the construction and operation of the Virgo detector
and the creation and support  of the EGO consortium. 
The authors also gratefully acknowledge research support from these agencies as well as by 
the Council of Scientific and Industrial Research of India, 
the Department of Science and Technology, India,
the Science \& Engineering Research Board (SERB), India,
the Ministry of Human Resource Development, India,
the Spanish Agencia Estatal de Investigaci\'on,
the Vicepresid\`encia i Conselleria d'Innovaci\'o, Recerca i Turisme and the Conselleria d'Educaci\'o i Universitat del Govern de les Illes Balears,
the Conselleria d'Innovaci\'o, Universitats, Ci\`encia i Societat Digital de la Generalitat Valenciana and
the CERCA Programme Generalitat de Catalunya, Spain,
the National Science Centre of Poland and the Foundation for Polish Science (FNP),
the Swiss National Science Foundation (SNSF),
the Russian Foundation for Basic Research, 
the Russian Science Foundation,
the European Commission,
the European Regional Development Funds (ERDF),
the Royal Society, 
the Scottish Funding Council, 
the Scottish Universities Physics Alliance, 
the Hungarian Scientific Research Fund (OTKA),
the French Lyon Institute of Origins (LIO),
the Belgian Fonds de la Recherche Scientifique (FRS-FNRS), 
Actions de Recherche Concertées (ARC) and
Fonds Wetenschappelijk Onderzoek – Vlaanderen (FWO), Belgium,
the Paris \^{I}le-de-France Region, 
the National Research, Development and Innovation Office Hungary (NKFIH), 
the National Research Foundation of Korea,
the Natural Science and Engineering Research Council Canada,
Canadian Foundation for Innovation (CFI),
the Brazilian Ministry of Science, Technology, and Innovations,
the International Center for Theoretical Physics South American Institute for Fundamental Research (ICTP-SAIFR), 
the Research Grants Council of Hong Kong,
the National Natural Science Foundation of China (NSFC),
the Leverhulme Trust, 
the Research Corporation, 
the Ministry of Science and Technology (MOST), Taiwan,
the United States Department of Energy,
and
the Kavli Foundation.
The authors gratefully acknowledge the support of the NSF, STFC, INFN and CNRS for provision of computational resources.
This work was supported by MEXT, JSPS Leading-edge Research Infrastructure Program, JSPS Grant-in-Aid for Specially Promoted Research 26000005, JSPS Grant-in-Aid for Scientific Research on Innovative Areas 2905: JP17H06358, JP17H06361 and JP17H06364, JSPS Core-to-Core Program A.\ Advanced Research Networks, JSPS Grant-in-Aid for Scientific Research (S) 17H06133, the joint research program of the Institute for Cosmic Ray Research, University of Tokyo, National Research Foundation (NRF) and Computing Infrastructure Project of KISTI-GSDC in Korea, Academia Sinica (AS), AS Grid Center (ASGC) and the Ministry of Science and Technology (MoST) in Taiwan under grants including AS-CDA-105-M06, Advanced Technology Center (ATC) of NAOJ, and Mechanical Engineering Center of KEK.

\appendix
\section{Tier-1 and Tier-2 Searches for Multiple Images}\label{app:strong_lensing_ll_pipelines_and_result}

In this Appendix, we present more details on the tier-1 methods used to analyze the whole set of \ac{GW} pairs considered in this work.

\subsection{Posterior Overlap}\label{app:posterior_overlap}

The \textsc{PO} approach~\citep{Haris:2018vmn}, already used in past lensing
searches~\citep{Hannuksela:2019kle,LIGOScientific:2021izm,LIGOScientific:2023bwz,Janquart:2023mvf}, directly looks at
the consistency between the posterior probability distributions obtained by traditional parameter-estimation analyses on the two events. It then
calculates the overlap between them, leading to a ranking statistic, denoted $B_{\mathrm{U}}^{\mathrm{L}}$.
In parallel, \ac{PO} also looks at the compatibility between the measured time delay and the one expected from galaxy lenses.
Requiring such compatibility has been shown to efficiently reduce the \ac{FPP}~\citep{Haris:2018vmn,Wierda:2021upe,More:2021kpb,Janquart:2022zdd}.
This second ranking statistic is noted $\mathcal{R}_{\mathrm{U}}^{\mathrm{L}}$. The final statistic for the search
is then obtained by multiplying the two. This product is then compared with the same statistic obtained from unlensed background events (see Appendix~\ref{app:background}), which is used to compute the \ac{FPP}.

\subsection{Phazap}\label{app:phazap}

\textsc{Phazap}~\citep{Ezquiaga:2023xfe} gauges the lensed nature of a system by looking for matching phases upon arrival at the detectors.  
The problem can be reduced to a six-dimensional parameter space, consisting of $\left\{ \phi_{\text{H}}, \phi_{\text{L}}, \phi_{\text{V}}, \tau_{\text{HL}}, \tau_{\text{HV}}, \Delta \phi_{f}  \right\}$
representing the phases at each detector, the arrival-time difference and the frequency evolution of the phase.
The phases $\phi_{\text{H,L,V}}$ of the signals under consideration are reconstructed at a reference frequency (40 Hz) for each of the three detectors,
while $\Delta\phi_f$, the difference in orbital phase for a given detector at two different frequency points, spans from 20 Hz to 100 Hz.
If a pair of signals is indeed strongly lensed, these six phasing parameters should coincide for the two signals, modulo a possible overall phase offset at the detector phases of $\pi/2$ or $\pi$ due to the Morse phase. In the presence of detector noise and measurement uncertainties, we have a distribution for each parameter instead of a point estimate which is derived directly from the posterior samples of each event.
We test their consistency with the strong-lensing hypothesis by measuring their distance in the six-dimensional phase space.
The compatibility between phases is then quantified by a $p$-value. A similar statistic is computed for the unlensed background events (see Appendix~\ref{app:background}), and the \ac{FPP} is then computed as the fraction of unlensed pairs with a $p$-value smaller than the one obtained for the lensed pair.

\subsection{Fast-GOLUM}\label{app:fastgolum}

\textsc{Fast-GOLUM}~\citep{Janquart:2021qov, Janquart:2023osz} uses the posterior of the first image, obtained by analyzing it under the strong-lensing hypothesis, to compute the conditioned evidence. This is done by performing nested sampling on the relative lensing parameters. By multiplying the conditioned evidence with the evidence of the first image, one obtains the joint evidence for strong lensing. Since this first image posterior is already concentrated on the region of interest in parameter space and one can further speed up calculation by using a look-up table, computing the joint evidence is accelerated. Dividing this joint evidence by the product of the unlensed evidences of the two images gives the coherence ratio
\begin{equation}
    \mathcal{C}^\mathrm{L}_\mathrm{U} = \frac{p(d_2 | d_1, \mathrm{L})p(d_1 | \mathrm{L})}{p(d_1 | \mathrm{U})p(d_2 | \mathrm{U})}
    \, ,
\end{equation}
where $d_i$ is the data for the $i^{\mathrm{th}}$ signal, and $\mathrm{L}$ and $\mathrm{U}$ denote the lensed and unlensed hypotheses, respectively. 

\begin{figure}[t]
    \centering
    \includegraphics[width=0.5\textwidth]{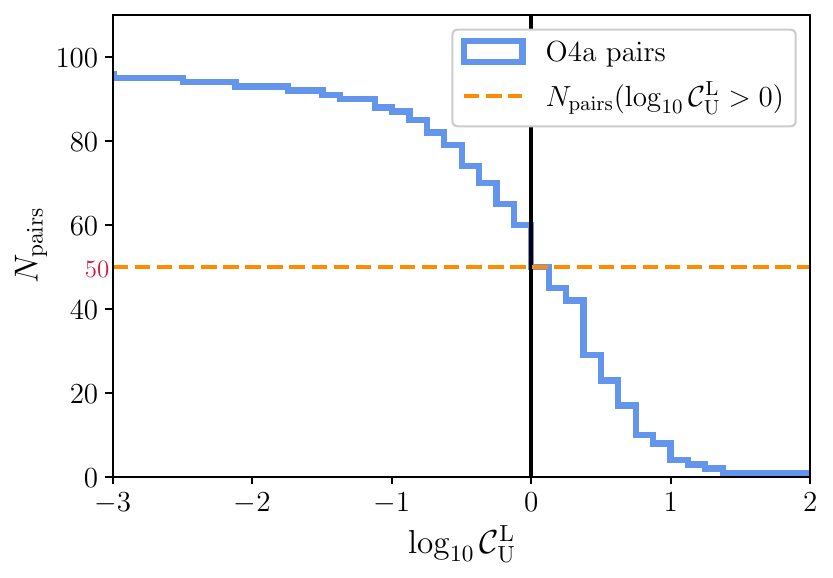}
    \caption{Cumulative histogram showing the distribution of the coherence ratio $\log_{10} \mathcal{C}^L_U$ of the lensed evidence against the unlensed evidence for the pairs passed to \textsc{Fast-GOLUM}. We only show pairs with $\log_{10} \mathcal{C}^L_U > -3$. This analysis keeps 50 pairs to be followed up by tier-3 analyses.}
    \label{hist_golum_im2}
\end{figure}

This approach is used for all event pairs flagged as interesting by tier-1 analyses. Figure~\ref{hist_golum_im2} shows the distribution of the coherence ratio $\log_{10} \mathcal{C}^\mathrm{L}_\mathrm{U}$ of the 105 pairs analyzed by \textsc{Fast-GOLUM}.

\section{Astrophysical Background}\label{app:background}

To construct our astrophysical background, we simulated around $40{,}000$ \ac{BBH} signals in the LIGO-only detector network, assuming the two detectors are operational for all detections and requiring network optimal \ac{SNR} $> 8$. From these, a random subset of $350$ signals was selected for further analysis. The mass and spin priors were derived from the inferred population model based on GWTC-3.0 data~\citep{KAGRA:2021duu}: the source-frame component masses were drawn from the \textsc{PowerLaw+Peak} distribution with a higher-mass cut extended at 200 $M_\odot$, and the spins were assumed to be independent and identically distributed, with spin magnitudes following a Beta distribution and spin orientations (tilts) modeled as an isotropic plus truncated half-Gaussian mixture~\citep{Talbot:2019okv}.
The merger rate density was assumed to follow the Madau--Dickinson profile~\citep{Madau:2014bja, Fishbach:2018edt}, with cosmological parameters consistent with~\cite{Planck:2015fie}.

\begin{figure}
	\centering
	\includegraphics[keepaspectratio, width=0.5\textwidth]{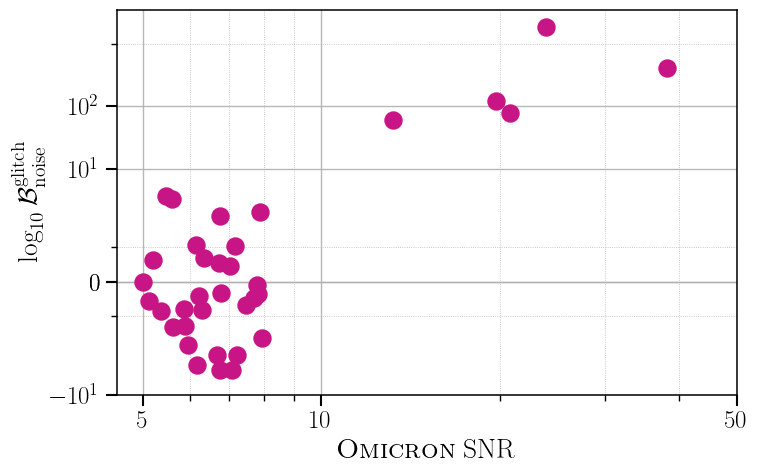}
	\caption{Comparison between the Bayes factor for the presence of a glitch versus purely Gaussian noise ($\log_{10}\mathcal{B}^{\rm glitch}_{\rm noise}$), computed using \textsc{BayesWave}, and the Omicron trigger \ac{SNR} across various data segments. For Omicron SNRs greater than $10$, the Bayes factor is consistently high ($>100$), indicating strong agreement between the two measures. However, for SNRs below $8$, there are instances where the Omicron SNR is low, yet the Bayes factor remains significant (greater than $1$), and in some cases even higher (exceeding $3$), which suggests strong disagreement between the two measures according to the Kass–Raftery scale~\citep{Kass:1995loi}.}
	\label{fig:BW_glitch_BF_vs_omicron}
\end{figure}

These \ac{BBH} signals were then injected into detector data from the \ac{O4a} observing run. To avoid loud glitches or the presence of other \ac{GW} signals, we scanned the detector data using the \textsc{Omicron} pipeline~\citep{Robinet:2020lbf} and excluded stretches with \ac{SNR} $> 6$. The remaining data was divided into 32~s segments and further cleaned for residual loud non-Gaussian features using  \textsc{BayesWave}~\citep{Pankow:2018qpo,Cornish:2020dwh,Chatziioannou:2021ezd,Hourihane:2022doe}.
This cleaning was required since glitches can persist in the data even where \textsc{Omicron} trigger has \ac{SNR} lower than $6$, as can be seen in Figure~\ref{fig:BW_glitch_BF_vs_omicron}. Here, cleaning is done before injections, which is done for computational efficiency, but could lead to an over-correction of non-Gaussian features.
The waveforms, generated using \textsc{IMRPhenomXPHM-SpinTaylor} waveform model~\citep{Pratten:2020ceb, Colleoni:2024knd},
were then added to these cleaned segments.
The \ac{PSD} estimated by \textsc{BayesWave} was used for computing likelihoods during parameter estimation.

\begin{figure}
	\centering
	\includegraphics[keepaspectratio, width=1\textwidth]{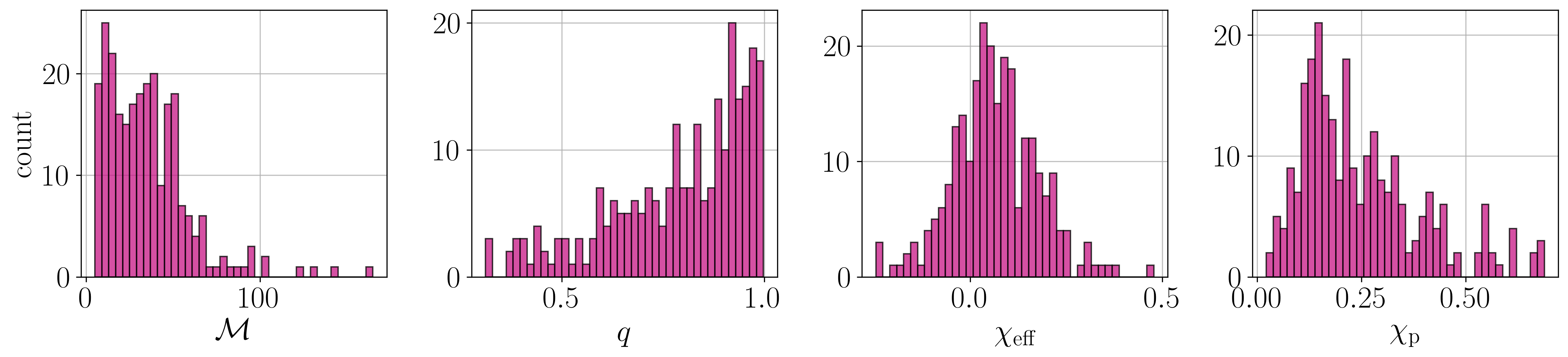}
	\caption{Distributions of the intrinsic parameters characterizing the background population: chirp mass $(\mathcal{M})$, mass ratio $(q)$, effective inspiral spin $(\chi_{\rm eff})$, and effective precession spin $(\chi_{\rm p})$.}
	\label{fig:bkg_pop_dist}
\end{figure}

Out of the $350$ injected and analyzed signals, $254$ had a recovered network matched-filter \ac{SNR} greater than $8$. Those were considered detections, or simulated observations, for our astrophysical background.
Figure~\ref{fig:bkg_pop_dist} shows the distribution of intrinsic parameters characterizing this background population.

\section{Additional Results for GW231123}\label{app:GW231123_extras}

\subsection{Posteriors Inferred From Point-Mass Analyses with Various Waveform Models}\label{app:GW231123_pointmass_posteriors}
\begin{figure*}
  \centering
  \includegraphics[keepaspectratio, width=0.495\textwidth]{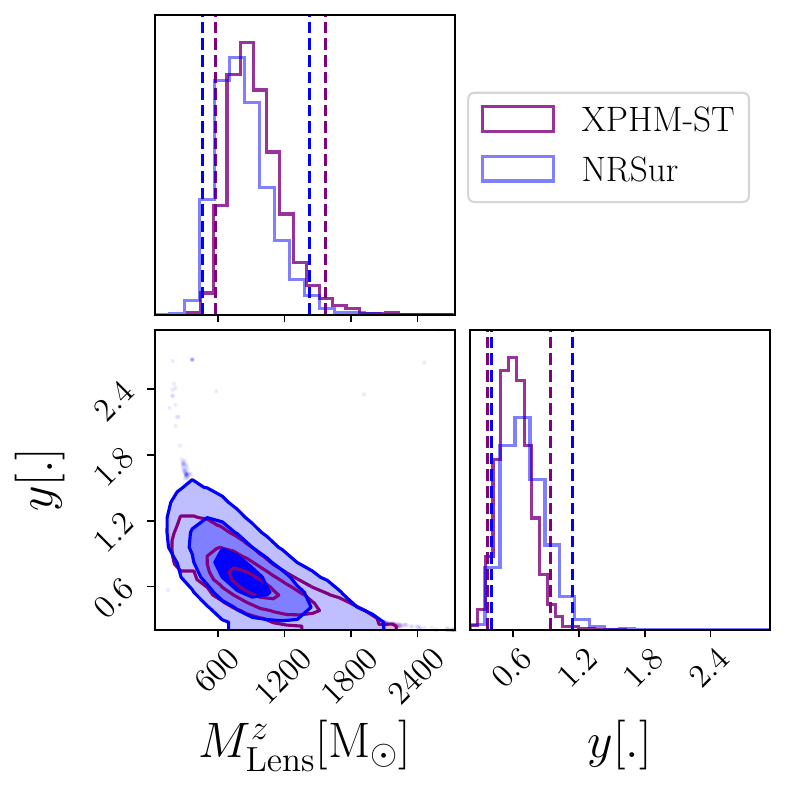}
  \includegraphics[keepaspectratio, width=0.495\textwidth]{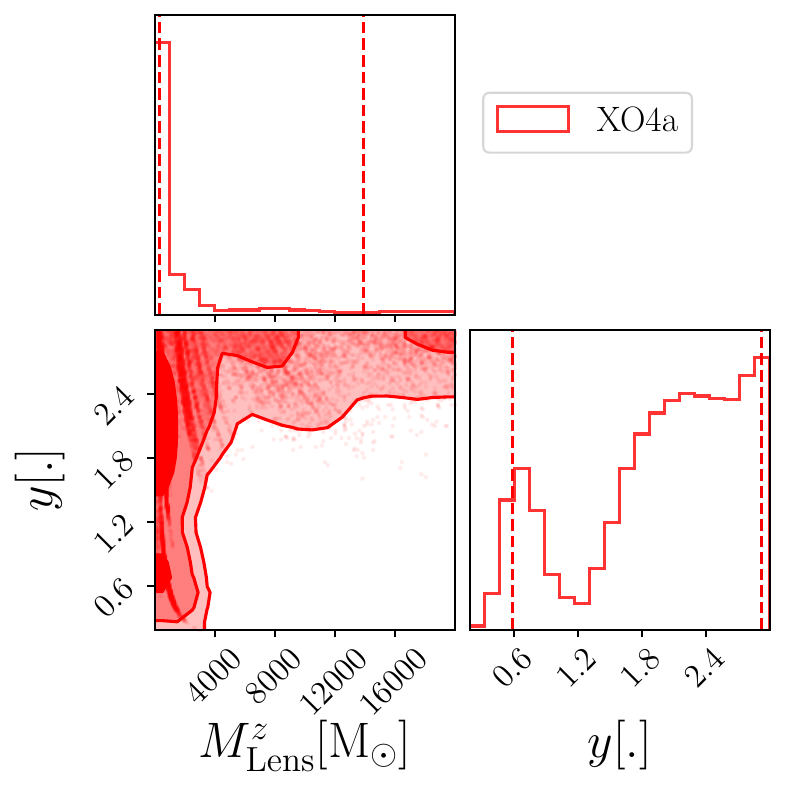}
  \caption{
    Posteriors of redshifted lens mass and dimensionless impact parameter inferred from the point-mass analyses with \textsc{IMRPhenomXPHM-SpinTaylor} (\textsc{XPHM-ST}; purple), \textsc{NRSur7dq4} (\textsc{NRSur}; blue), and \textsc{IMRPhenomXO4a} (\textsc{XO4a}; red) waveforms.
    The vertical dashed lines indicate the edges of the 90\% credible intervals.
    Consistent with the lensing supports shown in Table~\ref{tab:Bayes_factors_231123}, the \textsc{XPHM-ST} and \textsc{NRSur} analyses show well-converged posteriors with a clear peak, whereas the analysis with the \textsc{XO4a} yields broader posteriors, reflecting weaker lensing support.
  }
  \label{fig:GW231123_pointmass_posteriors}
\end{figure*}

The inferred lens properties from the point-mass analyses vary depending on the waveform model used.
Figure~\ref{fig:GW231123_pointmass_posteriors} shows the posteriors obtained from the point-mass analyses using three waveforms.

\subsection{Mismatch Between \textsc{cWB} and Waveform Models}\label{app:GW231123_waveform}

\begin{figure}
  \centering
  \includegraphics[keepaspectratio, width=0.9\textwidth]{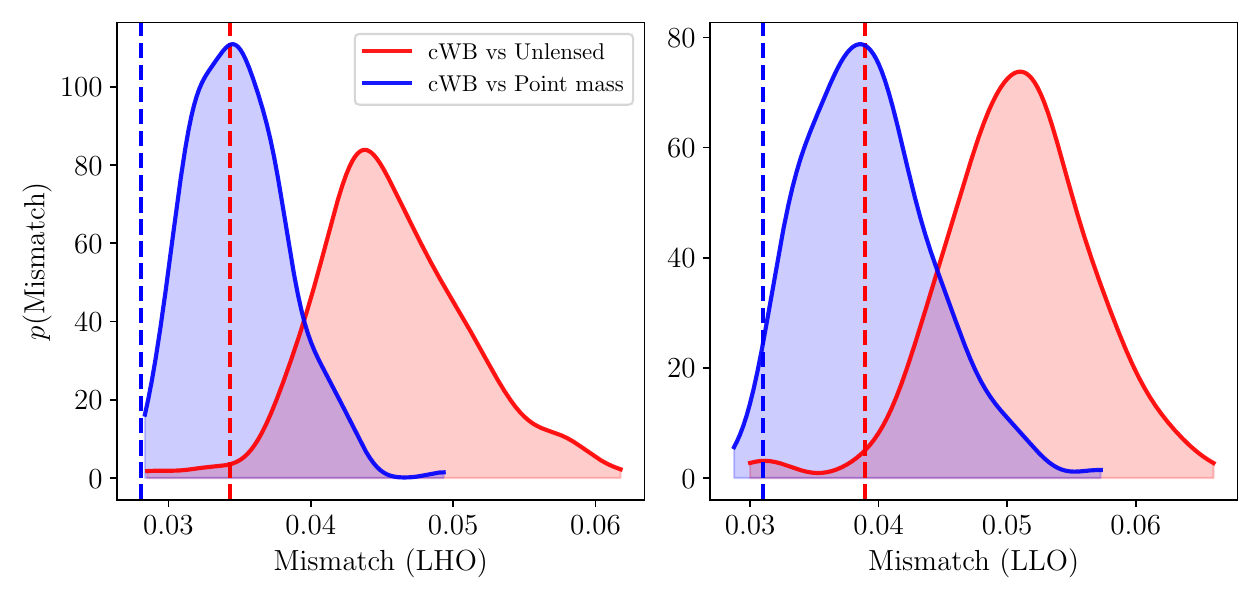}
  \caption{
    Distribution of the mismatch between the \textsc{cWB} reconstruction and the recovered lensed and unlensed distributions
    for all waveforms in the 90\% credible interval. The vertical dashed lines indicate the mismatch values between the \textsc{cWB} reconstruction
    and mean unlensed (red) and mean lensed (blue) waveforms, respectively.
    The lensed reconstructed signal generally better matches the model-agnostic reconstructed signal than the unlensed one.
  }
  \label{fig:GW231123_waveform_mismatch}
\end{figure}

In addition to the values reported for the mean waveforms in Section~\ref{subsec:cwb_comparison},
here we also show the distribution of mismatch values between the \textsc{cWB} reconstruction and the recovered lensed and unlensed distributions for all waveforms in the 90\% credible interval. 
Figure~\ref{fig:GW231123_waveform_mismatch} shows the distribution of mismatch values for each LIGO detectors. 
The mismatch values between the \textsc{cWB} reconstruction and the mean reconstructed unlensed and lensed waveforms are
$0.034$ ($0.039$) and $0.028$ ($0.031$) for {LHO} ({LLO}), respectively, 
indicating that the representative lensed reconstruction provides a closer match to the model-agnostic case. This is further shown by the median and 90\% credible intervals found for the mismatch distribution: $0.045^{+0.010}_{-0.007}$ and $0.051^{+0.010}_{-0.009}$ for LHO and LLO for the unlensed model, and $0.035^{+0.007}_{-0.005}$ and $0.039^{+0.010}_{-0.007}$ for the lensed case.
The difference between the distributions in the lensed and unlensed case is further quantified using the KS statistic, yielding values of 0.83 for LHO and 0.76 for LLO, indicating that, although the distributions appear relatively close, they are statistically different.
While such improvement shows that the lensed model is more consistent with the \textsc{cWB} reconstruction, it can be attributed to the additional degrees of freedom in the lensed model and is not a definitive proof of the lensed nature of the event. 

\clearpage

\bibliography{references}

\end{document}